\documentclass{article}
\usepackage[utf8]{inputenc}
\usepackage[margin=1in]{geometry}
\usepackage[colorlinks]{hyperref}
\usepackage{graphicx}
\usepackage[labelfont=bf]{caption}
\usepackage{amsmath} 
\usepackage[style=nature, sorting=none, giveninits=true]{biblatex}
\newcommand*\subtxt[1]{_{\textnormal{#1}}}
\DeclareRobustCommand\_{\ifmmode\expandafter\subtxt\else\textunderscore\fi}

\hypersetup{linkcolor = blue}

\usepackage{lineno}
\nolinenumbers
\title{\vspace{5cm}\textbf{Physics-Informed Deep Learning Characterizes Morphodynamics of Asian Soybean Rust Disease}}

\date{August 2021}

\begin{document}
\pagenumbering{gobble}

\maketitle
\begin{center}
    Henry Cavanagh\textsuperscript{1}, Andreas Mosbach\textsuperscript{2}, Gabriel Scalliet\textsuperscript{2}, Rob Lind\textsuperscript{3}, Robert G. Endres\textsuperscript{1}* 
    
*Corresponding author: r.endres@imperial.ac.uk 
\end{center}

\noindent \textsuperscript{1}Imperial College London, Centre for Integrative Systems Biology and Bioinformatics, London, UK, SW7 2BU; \textsuperscript{2}Syngenta Crop Protection AG, Schaffhauserstrasse 101, 4332 Stein, Switzerland; \textsuperscript{3}Syngenta International Research Centre, Jealott's Hill, Berkshire, UK, RG42 6EY

\newpage
\pagenumbering{arabic}  
\begin{abstract}
Medicines and agricultural biocides are often discovered using large phenotypic screens across hundreds of compounds, where visible effects of whole organisms are compared to gauge efficacy and possible modes of action. However, such analysis is often limited to human-defined and static features. Here, we introduce a novel framework that can characterize shape changes (morphodynamics) for cell-drug interactions directly from images, and use it to interpret perturbed development of \textit{Phakopsora pachyrhizi}, the Asian soybean rust crop pathogen. We describe population development over a 2D space of shapes (morphospace) using two models with condition-dependent parameters: a top-down Fokker-Planck model of diffusive development over Waddington-type landscapes, and a bottom-up model of tip growth. We discover a variety of landscapes, describing phenotype transitions during growth, and identify possible perturbations in the tip growth machinery that cause this variation. This demonstrates a widely-applicable integration of unsupervised learning and biophysical modeling. 
\end{abstract}

\section*{Introduction}

\begin{figure}
\centering
\includegraphics[width=1.\linewidth]{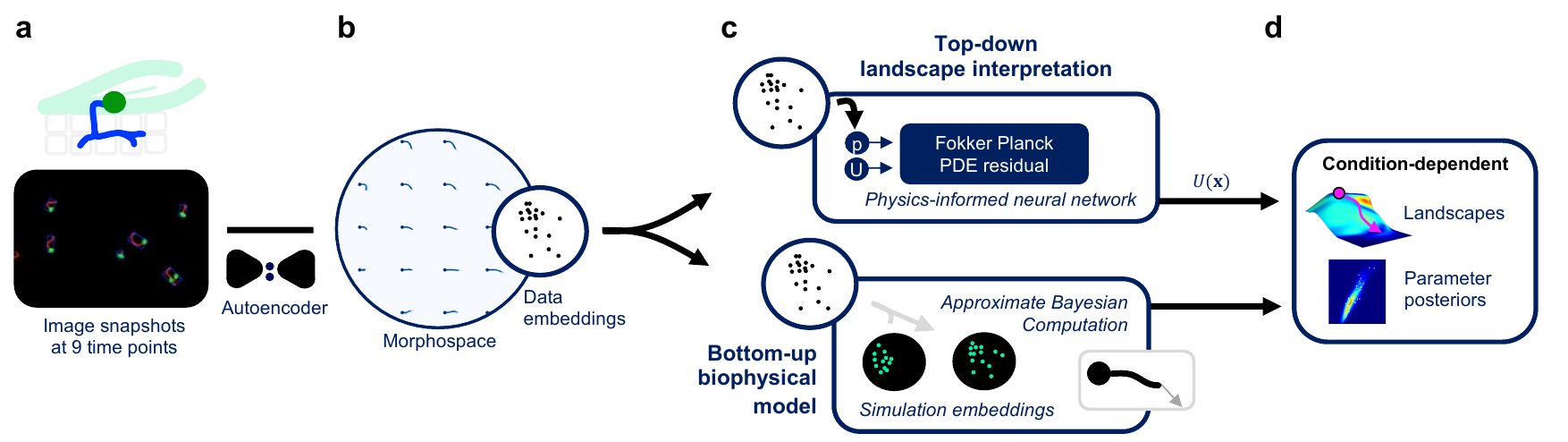}
\caption{\textbf{Morphodynamics of the Asian soybean rust pathogen, \textit{P. pachyrhizi}, are characterized through condition-dependent dynamics over a global morphospace.} (\textbf{a}) \textit{P. pachyrhizi} burrows into soybean leaves to extract nutrients, as sketched (top). Image sets at 9 time points under 6 conditions (bottom) are processed to yield aligned, single-fungus images. (\textbf{b}) An autoencoder learns the biophysical degrees of freedom from the images, discovering a 2D morphospace. (\textbf{c}) Dynamics are characterized using two models: a top-down landscape ($U(\mathbf{x})$) model, where a physics-informed neural network fits the Fokker-Planck equation to the morphospace embeddings, and a bottom-up persistent random walk model of the growth zone, with parameters fitted using approximate Bayesian computation with a morphospace-derived similarity metric. These yield (\textbf{d}) interpretable, condition-dependent characterizations in the form of Waddington-type landscapes and tip growth parameter posteriors.}
\label{fig:fig1}
\end{figure}

Quantifications of cell shape changes (morphodynamics) can reveal key developmental transitions and behavioural strategies, as well as modes of action of drugs by comparison with known drug-phenotype mappings \cite{nonejuie2013bacterial, mcdermott2021behavioral}. 
While recent progress in automated image analysis has popularized static descriptors beyond mean growth rates and metabolic fluxes \cite{usaj2016high}, the incorporation of dynamics can provide more complete system descriptions \cite{tweedy2019screening}, and may also aid the development and validation of mechanistic models \cite{keren2008mechanism}. Here, we developed such a framework and used it to interpret how fungicides affect the morphodynamics of \textit{Phakopsora pachyhizi}, the pathogen that causes rust disease in the soybean crop worldwide. Spores land on soybean leaves, grow germ tubes, penetrate the plant using appressoria, and subsequently form haustoria that extract nutrients \cite{fanaro2011asian}, as sketched in Fig. \ref{fig:fig1}a. This little-understood pathogen can cause economically devastating yield losses of up to 80 \% \cite{miles2003soybean} and is fast resistance-evolving \cite{langenbach2016fighting}.

Current methods for quantifying organism morphodynamics typically rely on a low-dimensional space of interpretable features \cite{berman2018measuring}. Though existing methods have uncovered remarkable behavioural patterns, revealing chemotactic strategies, temporal processing and social cooperation in a range of organisms \cite{tweedy2013distinct, berman2014mapping, brown2013dictionary, liu2018temporal}, typical shortcomings are as follows: first, they are often based on particular shape descriptors (e.g. 1D centerlines), restricting analysis to a narrow range of morphologies, often requiring sophisticated feature extraction algorithms. Second, they focus on stereotyped behaviours, which may not be characteristic of early development. Finally, states are often discretized and transition probabilities extracted, which obscures the continuous nature of morphodynamics. Alternatively, statistical analysis can be based on thousands of behavioral features to cluster and classify compounds and compare the results with known modes of action, but results can be difficult to interpret \cite{mcdermott2021behavioral}. In contrast to statistical correlates, interpretable continuous and stochastic models can be considered, but are not yet associated with morphodynamics. For instance, in Waddington's epigenetic landscape, cells begin at the top as pluripotent, and subsequently develop into a number of differentiated states, represented as lower-level valleys \cite{waddington2014strategy}. Despite many applications \cite{xu2014potential, huang2017processes, morris2014mathematical, wang2016geometrical, su2019phenotypic}, to our knowledge such landscapes have only been uncovered at steady-state.
    
A major method for image analysis, dimensionality reduction, and inference is deep learning \cite{angermueller2016deep}. Feedforward deep neural networks are comprized of a series of interconnected layers of artificial neurons, which transform inputs through a series of non-linear operations \cite{lecun2015deep}. Network parameters are updated through stochastic gradient descent and its variants in order to minimize a human-defined objective (the loss). These networks have been proven capable of approximating any function, given a sufficient number of parameters \cite{hornik1989multilayer}. To link data analysis and modeling, two types of network are particularly important: first, autoencoders, which are capable of capturing low dimensional structure in data through a constrained reconstruction task \cite{hinton2006reducing}. Due to their high flexibility, autoencoders have many advantages over non-parametric t-distributed stochastic neighbor embedding (t-SNE) and linear principal component analysis (PCA) algorithms. Second, physics-informed neural networks (PINNs). While forward problems can be solved fast with grid-based methods, PINNs can solve inverse (inference) problems for a wide range of differential equations \cite{raissi2019physics, raissi2020hidden}.

Here, we aimed to extract morphodynamic characterizations for intuitive comparison across compounds. Specifically, we analyzed \textit{P. pachyrhizi}, germinating \textit{in vitro} in a control compound and several fungicides, by first using an autoencoder to uncover a single 2D morphospace of salient features from high-throughput images of fixated fungi (Fig. \ref{fig:fig1}a-b). We then fitted two minimal models of dynamics over this morphospace and took their condition-dependent parameters as informative characterizations (Fig. \ref{fig:fig1}c-d). The two models approach the dynamics from opposite directions: the first is a top-down model that utilizes a Fokker-Planck description to uncover the global morphodynamic driving forces in the form of Waddington-type landscapes, and the second is a bottom-up persistent random walk model of the growth zone at the tip. We used a PINN to infer the landscapes, and approximate Bayesian computation, in conjunction with a morphospace-derived similarity metric, to infer the parameter posteriors of the tip growth model. The uncovered landscapes show that morphodynamics are diffusion-dominated until the germ tube begins to bend, at which point deterministic forces begin to drive trajectories apart. Fungicide-induced deformations include barriers, plateaus and canalized pathways, which may arise from differing stabilities of the growth zone. To avoid `black-box' methods, we analyzed our PINN to interpret the convergence. Our work shows how intuitive system characterizations can be acquired directly from images, by integrating unsupervised learning and biophysical modeling.

\section*{Results}
\subsection*{A Global Morphospace Reveals Perturbed Morphodynamics}

\begin{figure}[!ht]
\centering
\includegraphics[width=.5\linewidth]{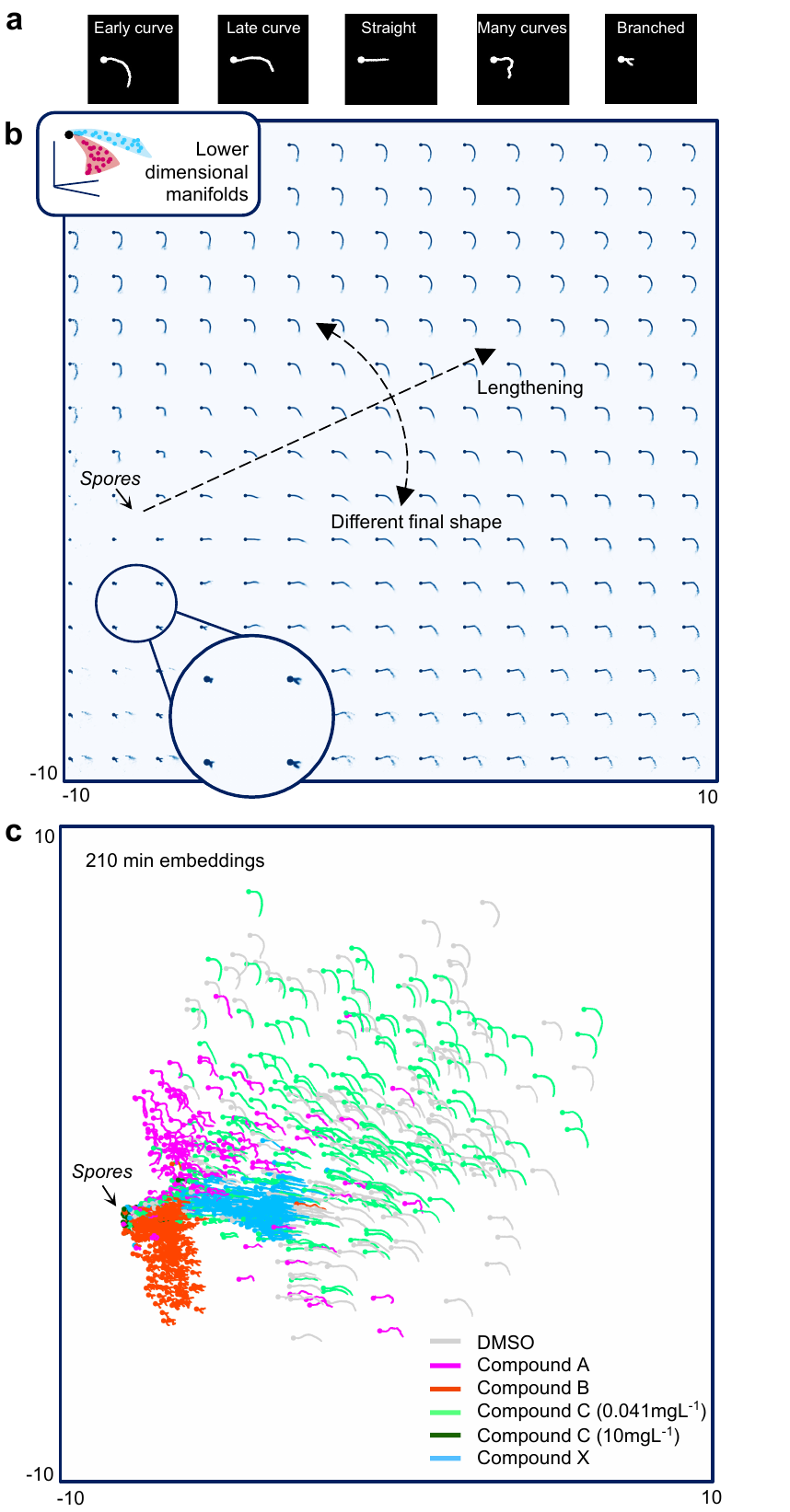}
\caption{\textbf{Global morphospace learned by a convolutional autoencoder.} (\textbf{a}) Human categorization of \textit{P. pachyrhizi} phenotypes is time consuming and introduces human biases and unnatural discretization. (\textbf{b}) A convolutional autoencoder addresses these shortcomings by learning the manifold associated with each condition, sketched inset. Morphospace features are shown by propagating morphospace coordinates on a grid through the decoder. (\textbf{c}) 210 min embeddings for all conditions show that fungicides can induce perturbed dynamics over the morphospace, which therefore represents for an expressive space for differentiating morphodynamics upon.}
\label{fig:fig2}
\end{figure}

Manual categorization of phenotypes is time consuming and limited to discrete human-defined features (Fig. \ref{fig:fig2}a). In contrast, manifold-based dimensionality reduction can provide a continuous low-dimensional space where dynamics are as simple as possible \cite{chan2020quantitative}. This is because an imaged dynamic system with $n$ degrees of freedom traces out an $n$-dimensional manifold within the higher dimensional pixel space, irrespective of the image dimensionality. 

To learn such a morphospace for \textit{P. pachyrhizi} growth, we carried out a high-throughput imaging assay of distinct populations at 9 equally-spaced time points, between 90 and 210 minutes after mixing with different compounds (code names used henceforth given in brackets): a dimethylsulphoxide control (DMSO), methyl benzimidazol-2-ylcarbamate at 1.1 mgL\textsuperscript{-1} (carbendazim, Compound A), PIK-75 hydrochloride at 3.3 mgL\textsuperscript{-1} (Compound B),  benzovindiflupyr (Compound C) at 0.041 mgL\textsuperscript{-1} and 10 mgL\textsuperscript{-1}, and Compound X (a Syngenta research compound related to trifluoromethyloxadiazoles \cite{winter2020trifluoromethyloxadiazoles}, see Supplementary Fig. 7 for the chemical structure) at 1.1 mgL\textsuperscript{-1}. These compounds were identified at pre-screening to show a range of phenotypes. We henceforth refer to each combination of compound and concentration as a condition. We extracted single-fungus images from the snapshot data sets, aligned such that the initial growth directions coincided, using automated processing, which yielded approximately 600,000 images in total (with mean and standard deviation across snapshots of approximately 11,000 and 3,000 respectively).  In order to validate the inferred Fokker-Planck model parameters and to motivate the tip growth model, we also gathered a small number of time-lapse videos of individual fungi for each compound (3-8, see Supplementary Movies 1-5), and aligned these manually. High-throughput analysis is not carried out with time-lapse videos in industry due to technical limitations. See Methods and Supplementary Note 1 for details on the compounds, imaging and image processing, and Supplementary Fig. 4a for an example time-lapse sequence.

The morphodynamics are perturbed differently in different conditions, and so the manifold traced out by a \textit{P. pachyrhizi} population is condition-dependent (as sketched in the inset of Fig. \ref{fig:fig2}b). To pull the condition-dependent manifolds together into a global morphospace, we trained a convolutional autoencoder (CAE) \cite{lecun1995convolutional}, an architecture specialized for images, with a 2D code space on images from all conditions and times (see Supplementary Note 1 for details on the architecture and training). To ensure the CAE used the morphodynamic manifolds to solve this reconstruction task, we selected the network complexity and training time that provided the simplest trajectories for embeddings associated with a small number of single-fungus videos. 

Figure \ref{fig:fig2}b shows the distribution of features over the morphospace, found by propagating morphospace coordinates on a grid through the decoder. Distance from the spore embedding loosely captures lengthening, and angle captures variation in final shape. Figure \ref{fig:fig2}c shows the 210 min embeddings for all conditions, revealing how fungicides can perturb dynamics over the morphospace. Some introduce novel features, e.g. the branching by Compound B, whereas others change the distribution over wild-type features, e.g. the increased prevalance of straightening induced by Compound X. In order to move from coarse-grained qualitative insights to quantitative characterizations, we next fitted two simple models of dynamics over the morphospace.

\subsection*{Emergent Landscapes Show the Morphodynamic Driving Forces}

\begin{figure}[ht]
\centering
\includegraphics[width=1.\linewidth]{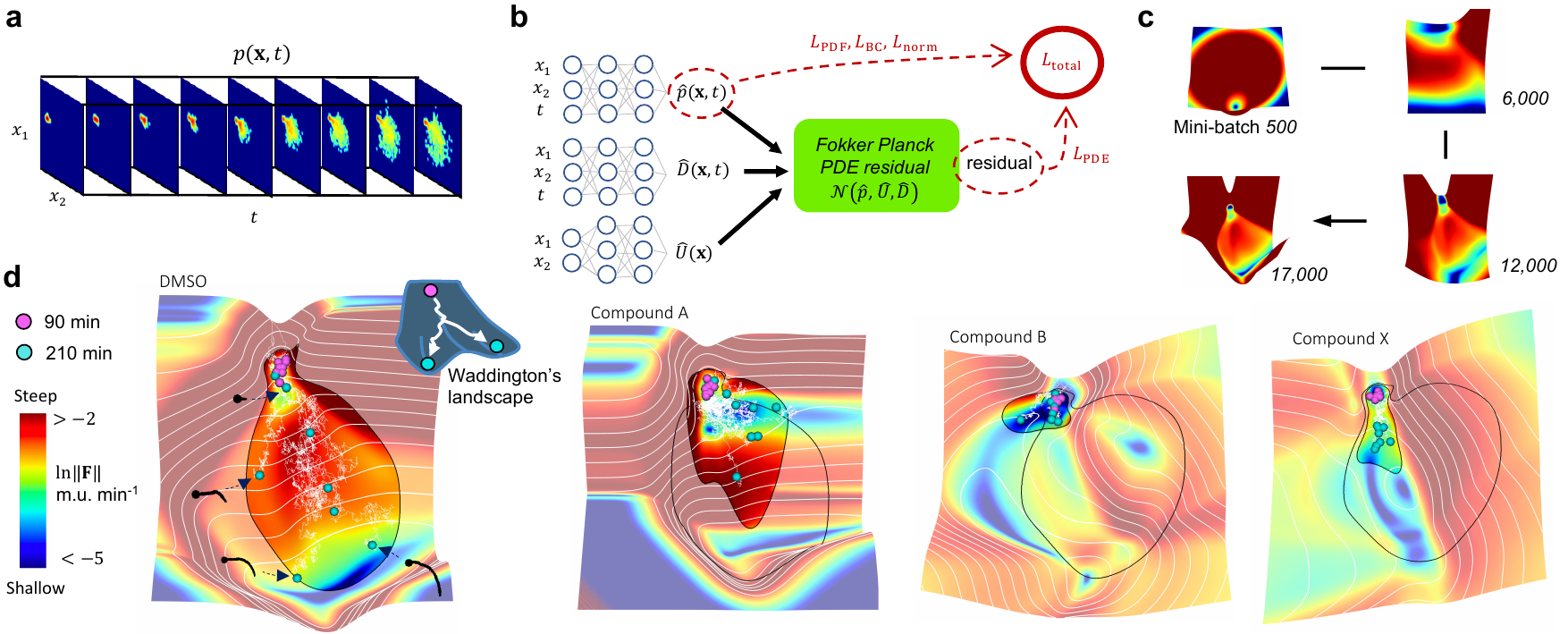}
\caption{\textbf{Morphodynamic landscapes learned by the PINN.} (\textbf{a}) Morphospace embeddings are transformed into probability density functions (PDFs), $p(\mathbf{x}, t)$, using kernel density estimation (KDE), yielding nine snapshots per condition. (\textbf{b}) A physics-informed neural network (PINN) learns the landscapes by fitting the Fokker-Planck equation to the PDFs. For each condition, the architecture comprizes a neural network to learn each of the PDF, $\hat{p}(\mathbf{x}, t)$, diffusivity, $\hat{D}(\mathbf{x}, t)$, and landscape, $\hat{U}(\mathbf{x})$. The outputs of these are put through a series of differential operators that outputs the Fokker-Planck residual, $\mathcal{N}$, and the architecture is trained to match the data ($L\_{PDF}$), minimize the magnitude of the residual ($L\_{PDE}$), satisfy the boundary conditions ($L\_{BC}$) and learn a normalized PDF ($L\_{norm}$).  (\textbf{c}) The architecture is trained over a series of mini-batches, with lower frequency solutions explored first. (\textbf{d}) Landscapes with simulated particles, from 90 mins (pink) to 210 mins (blue) after mixing with compounds, are shown, colored by the gradient magnitude, $\| \mathbf{F}\|$ (in terms of morphospace units, m.u.). These are analogous to Waddington's epigenetic landscape, as sketched in the inset. The black outlines show the contour where the PDF learned by the PINN is $10^{-3}$ for DMSO and each condition. The inner region therefore highlights areas with high data density, with the remaining areas shown to facilitate connection with the morphospace and outer tendencies. Contour lines are plotted along equal landscape values, with spacings of 0.11, 0.08, 0.07 and 0.09 m.u.\textsuperscript{2} min\textsuperscript{-1} for the landscapes from left to right. Morphodynamics are diffusion-dominated until the germ tube begins to bend, at which point deterministic forces begin to drive trajectories apart. Fungicide-induced deformations including barriers, plateaus and canalized pathways. This susceptibility to deformation, combined with the generality of the model, make the Fokker-Planck model well-suited for system characterization.}
\label{fig:fig3}
\end{figure}

Condition-dependent Waddington-type landscapes are intuitive  morphodynamic characterizations that can be inferred from the snapshot embeddings evolving over the morphospace. We model each condition as inducing a time-independent field of driving forces, $\mathbf{F}(\mathbf{x}; c, \boldsymbol{\lambda})$, that depends on the compound concentration, $c$, and any pharmacophores, captured in a parameter vector, $\boldsymbol{\lambda}$. In the absence of any curl, as we assume for early development, the force field can be associated with the gradient of a quasi-potential, $U$, via $\mathbf{F} = -\nabla U$, which we take as the developmental landscape. For cases where the underlying force field does have curl, landscapes can still be uncovered by splitting the force into curl-free and curl-containing components, yielding a  `potential and flux' description \cite{xu2014potential}.

To enable the inference of landscapes, we connected the evolving snapshot embeddings to the driving forces through a Fokker-Planck model. These embeddings were transformed into probability density functions (PDFs) on a grid using kernel density estimation (KDE, Fig. \ref{fig:fig3}a) \cite{davis2011remarks}. The Fokker-Planck partial differential equation (PDE) is used to separate out a system's driving forces and stochastic processes, and model statistical ensembles of Brownian particles. Each particle moves over the landscape according to the following stochastic differential equation:
\begin{equation}
    \mathbf{dx} = -\nabla U(\mathbf{x})dt + \boldsymbol{\sigma}(\mathbf{x}, t)\mathbf{dW},
\label{eq:sde}
\end{equation}
where $\mathbf{x}$ and t are the position in 2D space and time,  $ \boldsymbol{\sigma}(\mathbf{x}, t)$ is a noise matrix and $\mathbf{dW}$ is the Wiener process. $\boldsymbol{\sigma}(\mathbf{x}, t)$ has diagonal entries $\sqrt{2D(\mathbf{x}, t)}$ and zeros elsewhere, with $D$ being the diffusivity. The Fokker-Planck equation for the evolution of the PDF of the particle positions, $p(\mathbf{x}, t)$, is then 
\begin{equation}
    \frac{\partial p(\mathbf{x}, t)}{\partial t} = \sum_{i=1}^{2}\frac{\partial}{\partial x_{i}}\left[\frac{\partial U(\mathbf{x})}{\partial x_{i}}p(\mathbf{x}, t) + \frac{\partial}{\partial x_{i}}[D(\mathbf{x}, t)p(\mathbf{x}, t)]\right].
\end{equation}

To learn the Fokker-Planck landscapes (and diffusivities) given the PDF data, i.e. solve the inverse problem, we used a physics-informed neural network (PINN) \cite{raissi2019physics}. These learn PDE solutions by optimally matching PDF data, minimising the magnitude of the PDE residual, and satisfying any further constraints, e.g. boundary conditions. PINNs have several favorable properties over alternative methods for solving the inverse problem \cite{raissi2020hidden}. First, in going from the forward to the inverse problem, the only change required is the addition of extra learnable parameters; second, they can infer the governing equation with only sparse data, since they solve the inference of the full PDF and governing equation as a joint task \cite{chen2020solving}; third, they learn a continuous fully-differentiable solution, which means other variables of interest can be calculated directly from the learned variables, without numerical approximation (useful when transitioning between potential and force, for example); fourth, they learn progressively more complex functions as training progresses. As we show, this is a useful property when combined with early stopping if the required function complexity is not known \textit{a priori}. Finally, they scale more favorably with system dimensionality than grid-based methods, which can often perform well only for low-dimensional problems. This final property will prove especially useful when extending this work to higher-dimensional morphospaces.

We used one network to learn each of the PDF, diffusivity and potential (Fig. \ref{fig:fig3}b). The outputs of these ($\hat{p}(\mathbf{x},t)$, $\hat{D}(\mathbf{x},t)$ and $\hat{U}(\mathbf{x})$ respectively) are put through a series of differential operators ($\mathcal{N}$) that outputs the Fokker-Planck residual, 
\begin{equation}
\mathcal{N}(\hat{p}, \hat{U}, \hat{D}) = -\frac{\partial \hat{p}}{\partial t} + \sum_{i=1}^{2}\frac{\partial}{\partial x_{i}}\left[\frac{\partial \hat{U}}{\partial x_{i}}\hat{p} + \frac{\partial}{\partial x_{i}}(\hat{D}\hat{p})\right],
\label{eq:residual}
\end{equation}
which is incorporated into the loss so that the solution obeys the PDE.

The total loss to be minimized ($L\_{total}$) is the sum of four terms (shown in full in Methods). The first three are calculated over randomized mini-batches. They are the mean squared error between the data and learned PDF ($L\_{PDF}$), the mean squared PDF at the boundary ($L\_{BC}$), and the mean squared PDE residual ($L\_{PDE}$). The final term ensures the PDF is normalized, and is the squared difference between unity and a numerical integration over the full grid at a randomly-selected time point ($L\_{norm}$). The relative importance of these terms is determined by hyperparameters $a$, $b$, $c$ and $d$,
\begin{equation}
\label{eq:ltotal}
    L\_{total} = aL\_{PDF} + bL\_{BC} + cL\_{PDE} + dL\_{norm},
\end{equation}
and lower frequency functions are explored first (Fig. \ref{fig:fig3}c, \cite{rahaman2019spectral}), in alignment with Occam's razor. An ablation analysis also confirms that both space and time-dependent diffusion are required for the best model fit (Supplementary Fig. 1a).

Letting the PINN train to convergence would result in overfitting. This is the phenomenon where a neural network's function pushes beyond the problem-dependent desired complexity; for example, in image classification the network begins to learn spurious patterns and to generalize poorly to new data. In the context of inference from independent snapshots, overfitting corresponds to fitting to differences between individual snapshots that arise from variability between \textit{P. pachyrhizi} batches. While having independent snapshots is beneficial in that it shows a wider range of dynamics, this can result in features like a non monotonically-decreasing fraction of spores, which should not be captured in the model. The PINN learns trends common to all snapshots first, and we stop training before overfitting begins, identified by monitoring the spore region (as shown in Supplementary Fig. 1b-c for three training repeats). This regularization technique is known as early stopping. We note that the loss exponentially decays after an initial period of fast improvement. Hence, reasonable results can be achieved using much earlier stopping points than used here, if computational time is limited. Further details on the PINN architecture, hyperparameters and optimization can be found in Methods and Supplementary Note 2.

Figure \ref{fig:fig3}d shows the landscapes learned by the PINN, as well as a sample of forward simulations of Eq. \ref{eq:sde} (see Methods for details on the forward simulations), and the correspondence between the landscapes and morphospace is shown in Supplementary Fig. 2. The overfitting point described above approximately corresponds to 30, 30, 30, 20, 25 and 25 hours of training for DMSO and Compounds A, B, C (0.041 mgL\textsuperscript{-1}), C (10 mgL\textsuperscript{-1}) and X, respectively. Diffusion over the landscapes has two sources: morphodynamic diffusion, i.e. the fundamental unpredictability of morphology change from one time step to the next on the 2D morphological data manifold; and embedding noise, which arises because variations in image resolution, segmentation and alignment induce perturbations away from the manifold, and the complexity of the autoencoder's embedding function means these perturbations are not always just mapped to the closest point on the 2D manifold. Since the same image pre-processing algorithms were used on all conditions, differences in the Fokker-Planck diffusion over the same region of morphospace will be morphodynamic in nature, rather than arising from the embedding noise.

The condition-dependent landscapes are highly interpretable (Fig. \ref{fig:fig3}d). DMSO development begins in a metastable region of spores where diffusion dominates, with a threshold energy required for germination, and annealing diffusivity capturing a subpopulation of spores that never cross this threshold (Supplementary Fig. 2a). This may be similar to germination mechanics in fission yeast, where a polar cap stochastically wanders as the spore grows and ultimately breaks out once a critical strain is passed \cite{bonazzi2014symmetry}. Morphodynamics are diffusion-dominated until the germ tube begins to bend, at which point deterministic forces begin to drive trajectories apart. Compound A introduces a plateau in the region where tip bending begins, revealing growth stunting, and also opens up new features with numerous bends. Extreme cases of heteregeneous growth, for example the difference between fungi with dramatically slowed growth and others with completely unhindered growth, are captured by the Fokker-Planck model through plateaus followed by abnormally high gradients, shown dark red. Compound B opens up a new branching feature region immediately following germination. Some fungi do develop normally, but without significant bending, and with reduced growth rates. Compound X canalizes trajectories through only a subset of the features expressed in DMSO, namely straightening, with fungi closely following the landscape, before stunting occurs at a common location for most. Supplementary Fig. 1-2 show that Compound C at 0.041 mgL\textsuperscript{-1} inhibits growth slightly, primarily through reduced diffusion rather than reduced landscape gradients, and increasing the concentration to 10 mgL\textsuperscript{-1} drastically increases the stability of the spore region. 

The data PDFs compare well with PDFs generated by a kernel density estimate (KDE) of simulated trajectories for all conditions, validating the solution accuracy (Supplementary Fig. 3). Images from the videos can also be embedded in the morphospace, and the mean squared displacements of the video trajectories with those of the simulations show good agreement (Supplementary Fig. 4b-c). Furthermore, the entropy of the simulations always increases (Supplementary Fig. 4d). To get a measure of the uncertainty in the inferred landscapes and diffusivities, we trained the PINN three times for each compound, which exposed the algorithm to different data mini-batches. For each condition, the total losses decreased at similar rates (meaning the solutions are equally good at each training time), with overfitting occurring at approximately the same point. We therefore quantified uncertainty through the standard deviation of the fields across repeats at the early stopping times described above (Supplementary Fig. 1b-f).  The landscapes therefore provide intuitive comparisons of morphodynamics across conditions.

\subsection*{Landscape Deformations are Caused by Perturbations in the Tip Growth Machinery}

\begin{figure}[!ht]
\centering
\includegraphics[width=.5\linewidth]{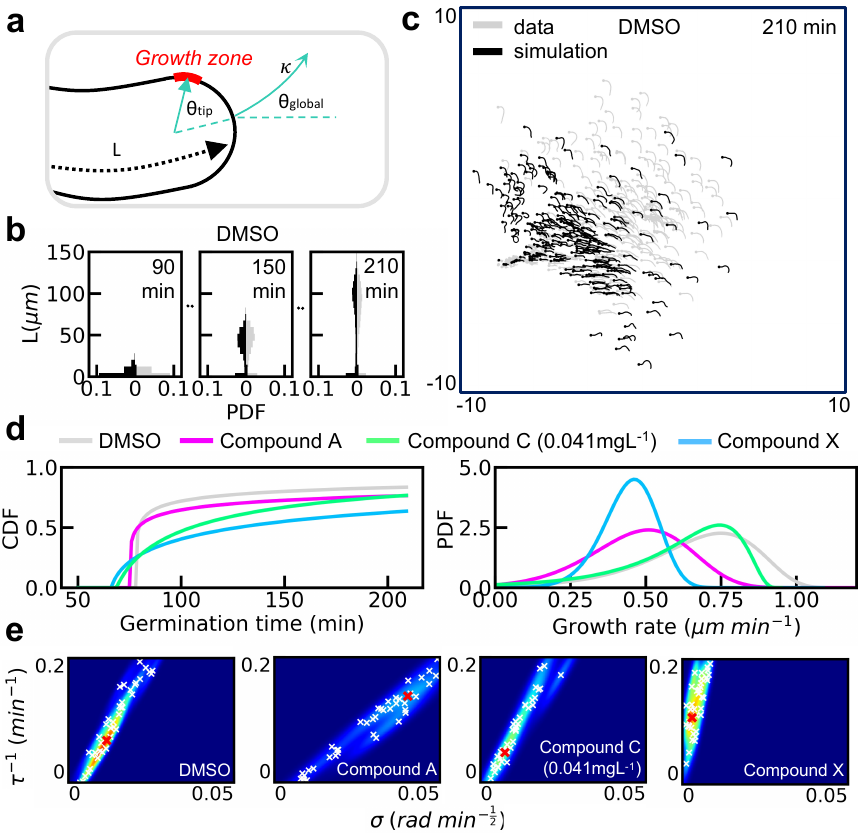}
\caption{\textbf{A persistent random walk model of the growth zone is fitted to image data.} (\textbf{a}) Tip growth is described with variables for length, $L$, linearly increasing in time, and path curvature, $\kappa$, which undergoes a persistent random walk, with relaxation to straight growth (i.e. a central growth zone). Dynamics of $\kappa$ may be the result of a diffusing growth zone, shown with angular location $\theta\_{tip}$, which causes a changing direction of growth, described by $\theta\_{global}$ in the lab frame. (\textbf{b}) All parameters were fitted using approximate Bayesian computation with sequential Monte Carlo (ABC-SMC). Lengthening parameters were fitted using length ($L$) histograms at 9 equally-spaced time points (3 of the 9 DMSO snapshots are shown here, with data in grey and simulations in black), between 90 and 210 min after mixing with solution. (\textbf{c}) Bending parameters were fitted by comparing morphospace embeddings of the 210 min snapshot data with those of images simulated with MAP lengthening parameters, as shown here for DMSO. (\textbf{d}) MAP germination time cumulative density functions (CDFs) and growth rate probability density functions (PDFs) show typical perturbations include premature germination, reduced germination frequency, and reduced maximum and mean growth rates. (\textbf{e}) Bending parameter posteriors (for stochasticity, $\sigma$, and relaxation to straight growth, $\tau^{-1}$) show final morphologies depend primarily on the ratio of the two bending parameters, and fungicides can both increase and decrease this ratio. Accepted parameters of the final ABC-SMC population are plotted in white, with MAP values in red. Source data are provided for (\textbf{d}-\textbf{e}).}
\label{fig:fig4}
\end{figure}

The morphospace can also be used for data-driven development of a minimal mechanistic model to reveal potential causes of the landscape deformations discovered above. Having a single model with condition-dependent parameters provides further cell-mechanical characterization. We developed such a model for tip growth under all but Compound B and Compound C at 10 mgL\textsuperscript{-1} (we excluded these because Compound B induces branching, which cannot be captured under the proposed model, and Compound C at 10 mgL\textsuperscript{-1} does not permit a significant germ tube to develop). The morphospace reveals final shape and associated lengthening to be the primary degrees of freedom across the populations, motivating equations for length, $L$, and tip bending.  From the videos, we observed three core features of lengthening (Supplementary Fig. 5a): that germination time is variable, that length increases approximately linearly with time, and that this growth rate is variable. We therefore modeled germination time, $t\_{g}$, and subsequent linear growth rate, $\alpha$, as lognormally-distributed, with growth rate distributed according to a reversed lognormal distribution truncated at zero. The lognormal distribution is widely used to model skewed phenomena across biology, including heterogeneous sensitivity to fungicides \cite{limpert2001log}. We compared three models for tip bending: a simple random walk in growth direction, $\theta\_{global}$ (Fig. \ref{fig:fig4}a); a random walk in the curvature of the growth path, $\kappa$; and a persistent random walk in $\kappa$ with parameters for stochasticity, $\sigma$, and relaxation to straight growth, $\tau^{-1}$ (see Methods for the mathematical expression of each model). Dynamics in $\kappa$ are a simple way to capture the effects of a diffusing growth zone, as has been described in fission yeast, where it is directed by landmark proteins concentrated at microtubule end points \cite{drake2013model}. This connection can be made slightly more explicit by the introduction of an angular growth zone position, $\theta\_{tip}$, and a mapping $\kappa = f(\theta\_{tip})$. While $f$ is unknown, this function is likely monotonically increasing, and passing through the origin (i.e. a central growth zone corresponds to straight growth).  Supplementary Fig. 5b shows $\theta\_{global}$ variation in time for the three models, for some intuition on the dynamics.

For both parameter inference and model selection, we used approximate Bayesian computation with sequential Monte Carlo (ABC-SMC, details on the SMC are in Methods) \cite{abcsmc}. In ABC, parameters are selected from a prior distribution and simulations are run. If the simulations are within some threshold of similarity with the data, then the parameters are stored. The density over the stored parameters forms the posterior distribution. For model selection, model index is introduced as an additional parameter and the posterior distribution is found over the joint space of model index and parameters, with the marginal distribution over model index giving the model probabilities. This biases towards low-dimensional parameter spaces, favoring the most parsimonious models. For the parameters involved in lengthening, we used histograms of fungus lengths at each snapshot (see Fig. \ref{fig:fig4}b for three of the nine DMSO snapshots). For the bending parameters, we then simulated fungus images using maximum \textit{a posteriori} probability (MAP) lengthening parameters, and compared histograms of their morphospace embeddings with those of the 210 min snapshot data (Fig. \ref{fig:fig4}c for DMSO). Further details can be found in Methods. Compound A data was used for model selection, as it covered the full spectrum of features. Only Model 3 could reproduce the data with high accuracy (in particular the multiple bends, where relaxation to central growth is required to reproduce the alternating bending direction). This match is shown by the probabilities of the three models and MAP simulations (Supplementary Fig. 5c-d). The equations governing dynamics after germination are therefore 
\begin{equation}
    \text{for} \; t > t\_{g} \begin{cases} dL = \alpha dt \\ d\kappa =  -\tau^{-1}\kappa dt + \sigma dW \end{cases},
\end{equation}
where dW is the Wiener process, and for $t \leq t\_{g}$ we have $L=\kappa=0$.

Figure \ref{fig:fig4}d shows the cumulative density functions (CDFs) of germination time and PDFs of growth rate associated with MAP parameters for each condition. Germination time is strongly skewed, with fungicides inducing premature germination and reducing subsequent germination frequency. Growth rates are less skewed, and both the maximum and mean growth rates are reduced by all fungicides.  The bending posterior distributions (Fig. \ref{fig:fig4}e) have linear shape, showing that for all conditions, the morphology depends primarily on the ratio of $\tau^{-1}$ to $\sigma$. Compounds A and X induce decreased and increased ratios of relaxation to stochasticity, respectively. Comparisons of MAP simulations with data for all conditions are shown in full in Supplementary Fig. 6.

\section*{Discussion}

Phenotypic screens are often used to identify drug efficacy and mode of action by comparing visible features like morphology. However, such screens are often limited to human-defined and static features, and any incorporation of dynamics typically focuses on stereotyped behaviors. Here, we characterized morphodynamics of the Asian soybean rust crop pathogen, \textit{P. pachyrhizi}, germinating \textit{in vitro} in the presence of different fungicides, directly from image sets. We found that morphodynamics are diffusion-dominated until the tip begins to bend, at which point deterministic forces begin to drive trajectories apart. Fungicide-induced landscape deformations include barriers, plateaus and canalized pathways. These features may arise from physical perturbations including premature but lower-frequency germination, reduced growth rates, and both increased and reduced stabilities of the growth zone. The global morphospace therefore allowed us to extract meaningful morphodynamic parameters directly from images, revealing perturbed driving forces in the Fokker-Planck model, and providing a similarity metric for the tip growth model. For both models, the non-linear embedding affords crucial interpretability through visualization. Moreover, the two models give complementary views of the dynamics. Taking Compound X as an example, the landscape model reveals low diffusion following germination when compared with DMSO. Hence, once germinated with Compound X, fungi grow at very similar rates, with little bending, tightly following the underlying landscape. The tip growth model similarly shows a narrowed distribution of growth rates and reduced bending, and these observations are confirmed when reviewing the time-lapse videos after the analysis.

Despite many benefits, the analysis in its current form has limitations. For systems with higher-dimensional dynamics, a 2D morphospace may be unsuitable. For instance, we omitted a compound that induced blistering along the germ tube, which expanded the compound's morphospace beyond two dimensions. For such cases, the bottom-up mechanistic model parameter fitting can be done in higher dimensions, but at the cost of increased computation time and reduced interpretability.  For the Fokker-Planck model, higher-dimensional systems may still be characterized in terms of networks of attractors \cite{wang2016geometrical}, and the marginal dynamics of pairs of morphological degrees of freedom could still be visualized in 2D. Such characterizing of dynamics over non-linear representations may be most powerful when different conditions cover similar features, such that differences have a physical, rather than algorithmic, origin. As well as being able to characterize dynamics across multiple conditions, these simple models are useful for systems without \textit{a priori} established dynamics. 

Do the landscapes correspond to any physical quantities beyond representing a distilled statistical representation? The potentials represent the deterministic part of the motion, e.g. extension of the germ tube from turgor pressure and vesicle delivery (the potentials largely drive in the direction of increasing length). Furthermore, diffusion captures features that vary at single-fungus level, including the precise growth rate, and the direction of bending (potentially caused by a diffusing growth zone or noise \cite{drake2013model}). The observed phenotypes are certainly plausible based on putative modes of action of the drugs in terms of inhibiting microtubules, kinases, and gene expression (see Supplementary Note 4).

An interesting area for future work is to extend unsupervised morphodynamic analysis beyond minimal characterizations, to more detailed models. This could be achieved by joint learning of the underlying representation and equations of motion e.g. by minimizing the prediction error and complexity of the equations of motion. Such approaches have been shown capable of recovering physical laws in Cartesian coordinates from warped video footage \cite{udrescu2021symbolic}, and it would be interesting to extend this to complex biological systems, where the underlying laws are less clear. Modeling of cell growth in terms of generalized shape coordinates is an area of active research, with one promising model balancing dissipative, mechanical and active forces \cite{banerjee2016shape}. Another interesting area for future work is to better understand the connection between internal mechanics and morphodynamics. This could be achieved by joint modeling morphology and organelles (using flourescent markers), conditional on various pharmacophores. Interesting organelles and molecular processes may include secretory vesicles for membrane delivery, small GTPases for growth cone labelling, and motor and cytoskeletal proteins for transport and structure \cite{campas2009shape}. Interpretable representations for each could be found using non-linear dimensionality reduction, as done here for morphology \cite{johnson2017building}.

Although often hailed as the future of deep learning, use of unsupervised learning techniques within the natural sciences often stops at low-dimensional data visualizations. We hope the work presented here may stimulate further work leveraging the discovery power of unsupervised methods within interpretable physical models.

\section*{Methods}

Algorithms were run in Python, and packages used are detailed in Supplementary Methods. 

\subsection*{Imaging and Image Processing}

For the snapshot data, spores were mixed in each of the 6 treatment solutions and imaged in 96-well plates on the Opera QEHS running Opera Software 2.0 (EvoShell, Opera CHKN/QEHS Red Ver. 2.0.0.12017 Rev.: 89046, PerkinElmer Inc.). At 9 equally-spaced times between 90 and 210 min, cell walls were stained with Calcofluor White solution with KOH, which fixated the fungi. Each snapshot was therefore of a different batch of spores. The staining procedure enabled the collection of two images for each well and time point, using different excitation wavelengths, one showing the spores and another showing the germ tubes. For the time-lapse videos, images were taken at 3 min intervals on the JuLI Stage Real Time Cell History Recorder running JuLI Stage V. 2.0.1 and JuLI EDIT V. 1.0.0.0 (NanoEnTek Inc.), without staining, meaning only one image was collected for each well and time, showing the full fungi. All imaging was done at 10$\times$ magnification. Full details on the imaging can be found in Supplementary Information. The snapshots and time-lapse videos were processed differently, because a) the snapshots had spores and germ tubes separated which we took advantage of to automate alignment, and b) the small number of time-lapse videos meant we could manually align these for higher precision.

For processing the snapshot images, we used adaptive binarization (to account for lighting defects) to get two images for each view: one of germ tube contours, another of spore contours. Adding these together then gave an image with full fungi. Contours with an area above a threshold found by trial and error were removed as obvious overlapping fungi, and the remaining ones were cropped by finding the minimum bounding rectangle. These regions of interest (ROIs) were rotated to align with the pixel grid and padded so all were $200\times200$ pixels, to fit the largest fungi in the set. Incomplete fungi were also removed at the image borders. We then used a supervised convolutional network, trained on a sample of hand-labelled images, to remove contours that contained overlapping fungi. Remaining individual fungi were then translated and rotated so the initial growth directions coincided, and a flip was executed if the right-most point of the fungus was higher than the germination point (see Supplementary Note 1). Finally, we replaced all spores with identical circles, so as to prioritize modeling of the germ tube; the resulting morphospace point is then widened into a spore region through the kernel density estimation. 

For the time-lapse videos, we again binarized the frames (non-adaptive this time, as there were not significant lighting defects). We then found series of contours across frames whose centers of mass were closest, and manually looked through these series to find those that corresponded to tracking an individual fungus. We then manually aligned these so that the initial growth directions coincided, this time using ImageJ and Gimp. Before being inputted into the autoencoder, snapshot and time-lapse video pixels were asigned to a value in the set \{0, 1\}. Full details on the image processing can be found in Supplementary Note 1.

\subsection*{Neural Networks}

For the autoencoder's encoder, we used four convolutional layers with 16, 32, 64 and 16 feature maps, all with 3$\times$3 kernels, ReLU activations, batch normalization, and alternating stride sizes of 1 and 2 in PyTorch. The decoder's structure  mirrored the encoder's, but with transposed convolutions. We used a sigmoid output activation and binary cross entropy loss, over mini-batches of 50 images, and trained for 4 epochs using the Adam optimizer \cite{kingma2014adam} with a learning rate of $10^{-4}$, which took 2 hours with a Quadro RTX 6000 GPU card. Training was stopped at the point at which the trajectories of the single-fungus videos were least complex.

For the PINN, the loss function to be minimized comprizes four terms, with the first three calculated over random mini-batches of $N$ data points, and the final one over the full spatial grid of $M$ data points. The first is the mean squared difference between the learned PDF, $\hat{p}(\mathbf{x}^{j},t^{j})$, and data, $p(\mathbf{x}^{j},t^{j})$,
\begin{equation}
    L\_{PDF} = \frac{1}{N} \sum_{j=1}^{N}{[ \hat{p}(\mathbf{x}^{j},t^{j}) - p(\mathbf{x}^{j},t^{j}) ]}^{2},
\end{equation}
with $\{\mathbf{x}^{j},t^{j}\}$ in the nine snapshots. The second is the mean squared PDF at the boundary, 
\begin{equation}
    L\_{BC} = \frac{1}{N} \sum_{j=1}^{N}{[\hat{p}(\mathbf{x}^{j},t^{j}) ]}^{2}
\end{equation}
with $\{\mathbf{x}^{j},t^{j}\}$ selected from $10^{6}$ uniformly distributed boundary points, and the third term is the mean squared PDE residual ($\mathcal{N}$, given in Eq. \ref{eq:residual}),
\begin{equation}
    L\_{PDE} = \frac{1}{N} \sum_{j=1}^{N}{[ \mathcal{N}(\hat{p}(\mathbf{x}^{j},t^{j}), \hat{D}(\mathbf{x}^{j},t^{j}) , \hat{U}(\mathbf{x}^{j})) ]}^{2},
\end{equation}
with $\{\mathbf{x}^{j},t^{j}\}$ selected from $10^{6}$ points uniformly distributed over the whole domain. The final term ensures the PDF integrates to one:
\begin{equation}
    L\_{norm} =  \left[ \sum_{j=1}^{M}{\Delta x_{1}\Delta x_{2} \hat{p}(\mathbf{x}^{j},t)} -1  \right]^{2},
\end{equation}
with $\mathbf{x}^{j}$ covering the full spatial grid and $t$ randomly selected. For the total loss (Eq. \ref{eq:ltotal}), we used hyperparameters of 1, 1, 500, 0.01 for $a$, $b$, $c$ and $d$, with reasons discussed in Supplementary Note 2.

The three PINN neural networks had 5 fully connected layers, each with 50 neurons, with residual skip connections, and swish activations between layers. Output variables that share inputs (e.g. the PDF and diffusivity) can be outputted from a single neural network if they are likely to have similar features, for increased computational efficiency. We used the Adam optimizer \cite{kingma2014adam} with a learning rate of $5\times10^{-4}$, and batch sizes, $N$, of 8,000. To speed up training, the DMSO landscape was first trained for 10 hours, and PINNs for the other conditions were initialized with these weights (known as transfer learning). For forward simulations over the landscapes, particle starting positions were sampled from the initial PDF learned by the PINN, and then simulations were run by evaluating the potential and diffusivity on a $1000\times1000$ spatial grid, with 20 snapshots in time for the diffusivity, and simulating Eq. \ref{eq:sde} with a time step of 0.01 min.

\subsection*{Tip Growth Model}

The 3-parameter lognormal probability density function is given by
\begin{equation}
    f(x; s, \sigma^{2}, \text{loc}) = \frac{1}{\sigma\sqrt{2\pi}(x-\text{loc})}\exp{\frac{\log^{2}\left(\frac{x-\text{loc}}{s}\right)}{2\sigma^2}}
\end{equation}
where $\sigma$ is a shape parameter, $s$ is a scale parameter (also the median), and $\text{loc}$ is a location parameter (the lower bound). The 2-parameter distribution has $\text{loc}$ set to zero.

We modeled germination time, $t\_{g}$, as distributed according to $t\_{g} \sim lognormal(s_{t\_{g}}, \sigma_{t\_{g}}, \text{loc}_{t\_{g}})$, and growth rate, $\alpha$, as distributed according to $\alpha = \text{loc}_{\alpha}-x$, with $x \sim lognormal(s_{\alpha}, \sigma_{\alpha}, 0)$ and resampling for negative $\alpha$. Length data was extracted by summing the binarized fungus images, and both the lengthening and bending parameters were fitted using ABC-SMC \cite{abcsmc}. This is a computationally efficient implementation of ABC, identifying intermediate distributions over a series of populations, and gradually decreasing the acceptance threshold. All histograms were compared using the summed absolute distance, and we trained the autoencoder for an extra two epochs with simulations generated randomly from the prior distribution to get coverage of any novel features. 

We compared three models for tip bending, with Model 3 found to reproduce the data best. For all of the following, $\sigma$ is a noise parameter that was fitted, and $dW$ is the Wiener process. Fig. \ref{fig:fig4}a shows a schematic with the bending angles and curvature. Model 1 was a random walk in the global direction, $\theta\_{global}$, a simple model commonly used in the literature:
\begin{equation}
    d\theta\_{global} = \sigma dW.
\end{equation}
Model 2 was a random walk in the curvature of the growth path, $\kappa$, in order to connect to cell tip mechanics:
\begin{equation}
    d\kappa = \sigma dW.
\end{equation}
Model 3 was a persistent random walk in the curvature, with an additional parameter, $\tau^{-1}$, for relaxation to straight growth, motivated by work analysing fission yeast tip growth mechanics \cite{drake2013model}:
\begin{equation}
    d\kappa = -\tau^{-1}\kappa dt + \sigma dW.
\end{equation}
See Supplementary Note 3 for details on the creation of the simulation images, and settings used for running ABC-SMC.

\vspace{5mm}

\noindent \textbf{Acknowledgements} \\
We thank Suhail Islam for invaluable computational suppport. This work was funded by the Biotechnology and Biological Sciences Research Council (grant number BB/M011178/1) and Syngenta provided financial and technical support in the form of an iCASE studentship to H.C. \\

\noindent \textbf{Author Contributions} \\
H.C. and R.G.E. designed and H.C. performed the theoretical analysis; A.M. and G.S. contributed the reagants and performed the imaging; H.C. and R.L. did the image processing. All authors wrote the paper. \\

\noindent \textbf{Competing Interests} \\
The authors declare no competing interests. \\

\noindent \textbf{Data Availability} \\
The image data that support the findings of this study have been deposited at \url{http://cellimagelibrary.org/groups/54615}. The figure data generated in this study are provided in the Supplementary Information/Source Data file. \\

\noindent  \textbf{Code Availability} \\
The code used, along with a subset of images, are available at \url{https://github.com/hcbiophys/morphodynamics} \cite{Cavanagh2021Physics}.

\printbibliography

@article{winter2020trifluoromethyloxadiazoles,
  title={Trifluoromethyloxadiazoles: inhibitors of histone deacetylases for control of Asian soybean rust},
  author={Winter, Christian and Fehr, Marcus and Craig, Ian R and Grammenos, Wassilios and Wiebe, Christine and Terteryan-Seiser, Violeta and Rudolf, Georg and Mentzel, Tobias and Quintero Palomar, Maria Angelica},
  journal={Pest Management Science},
  volume={76},
  number={10},
  pages={3357--3368},
  year={2020},
  publisher={Wiley Online Library}
}

@article{nonejuie2013bacterial,
  title={Bacterial cytological profiling rapidly identifies the cellular pathways targeted by antibacterial molecules},
  author={Nonejuie, Poochit and Burkart, Michael and Pogliano, Kit and Pogliano, Joe},
  journal={Proceedings of the National Academy of Sciences},
  volume={110},
  number={40},
  pages={16169--16174},
  year={2013},
  publisher={National Acad Sciences}
}

@article{mcdermott2021behavioral,
  title={Behavioral fingerprints predict insecticide and anthelmintic mode of action},
  author={McDermott-Rouse, Adam and Minga, Eleni and Barlow, Ida and Feriani, Luigi and Harlow, Philippa H and Flemming, Anthony J and Brown, Andr{\'e} EX},
  journal={Molecular systems biology},
  volume={17},
  number={5},
  pages={e10267},
  year={2021}
}

@article{usaj2016high,
  title={High-content screening for quantitative cell biology},
  author={Usaj, Mojca Mattiazzi and Styles, Erin B and Verster, Adrian J and Friesen, Helena and Boone, Charles and Andrews, Brenda J},
  journal={Trends in cell biology},
  volume={26},
  number={8},
  pages={598--611},
  year={2016},
  publisher={Elsevier}
}

@article{chan2020quantitative,
  title={Quantitative comparison of principal component analysis and unsupervised deep learning using variational autoencoders for shape analysis of motile cells},
  author={Chan, Caleb K and Hadjitheodorou, Amalia and Tsai, Tony Y-C and Theriot, Julie A},
  journal={bioRxiv},
  year={2020},
  publisher={Cold Spring Harbor Laboratory}
}

@article{tweedy2019screening,
  title={Screening by changes in stereotypical behavior during cell motility},
  author={Tweedy, Luke and Witzel, Patrick and Heinrich, Doris and Insall, Robert H and Endres, Robert G},
  journal={Scientific reports},
  volume={9},
  number={1},
  pages={1--12},
  year={2019},
  publisher={Nature Publishing Group}
}

@article{keren2008mechanism,
  title={Mechanism of shape determination in motile cells},
  author={Keren, Kinneret and Pincus, Zachary and Allen, Greg M and Barnhart, Erin L and Marriott, Gerard and Mogilner, Alex and Theriot, Julie A},
  journal={Nature},
  volume={453},
  number={7194},
  pages={475--480},
  year={2008},
  publisher={Nature Publishing Group}
}

@article{fanaro2011asian,
  title={The Asian Soybean Rust in South America},
  author={Fanaro, Gustavo B and Villavicencio, Anna Lucia CH},
  journal={Soybean: Physiology and Biochemistry},
  pages={475},
  year={2011},
  publisher={BoD--Books on Demand}
}

@article{miles2003soybean,
  title={Soybean rust: Is the US soybean crop at risk},
  author={Miles, Monte R and Frederick, Reid D and Hartman, Glen L},
  journal={APS net},
  year={2003}
}

@article{langenbach2016fighting,
  title={Fighting Asian soybean rust},
  author={Langenbach, Caspar and Campe, Ruth and Beyer, Sebastian F and Mueller, Andr{\'e} N and Conrath, Uwe},
  journal={Frontiers in plant science},
  volume={7},
  pages={797},
  year={2016},
  publisher={Frontiers}
}

@article{berman2018measuring,
  title={Measuring behavior across scales},
  author={Berman, Gordon J},
  journal={BMC biology},
  volume={16},
  number={1},
  pages={23},
  year={2018},
  publisher={Springer}
}

@article{berman2014mapping,
  title={Mapping the stereotyped behaviour of freely moving fruit flies},
  author={Berman, Gordon J and Choi, Daniel M and Bialek, William and Shaevitz, Joshua W},
  journal={Journal of The Royal Society Interface},
  volume={11},
  number={99},
  pages={20140672},
  year={2014},
  publisher={The Royal Society}
}

@article{brown2013dictionary,
  title={A dictionary of behavioral motifs reveals clusters of genes affecting Caenorhabditis elegans locomotion},
  author={Brown, Andr{\'e} EX and Yemini, Eviatar I and Grundy, Laura J and Jucikas, Tadas and Schafer, William R},
  journal={Proceedings of the National Academy of Sciences},
  volume={110},
  number={2},
  pages={791--796},
  year={2013},
  publisher={National Acad Sciences}
}

@article{tweedy2013distinct,
  title={Distinct cell shapes determine accurate chemotaxis},
  author={Tweedy, Luke and Meier, B{\"o}rn and Stephan, J{\"u}rgen and Heinrich, Doris and Endres, Robert G},
  journal={Scientific reports},
  volume={3},
  pages={2606},
  year={2013},
  publisher={Nature Publishing Group}
}

@article{liu2018temporal,
  title={Temporal processing and context dependency in Caenorhabditis elegans response to mechanosensation},
  author={Liu, Mochi and Sharma, Anuj K and Shaevitz, Joshua W and Leifer, Andrew M},
  journal={Elife},
  volume={7},
  pages={e36419},
  year={2018},
  publisher={eLife Sciences Publications Limited}
}

@book{waddington2014strategy,
  title={The strategy of the genes},
  author={Waddington, Conrad Hal},
  year={2014},
  publisher={Routledge}
}

@article{xu2014potential,
  title={The potential and flux landscape theory of ecology},
  author={Xu, Li and Zhang, Feng and Zhang, Kun and Wang, Erkang and Wang, Jin},
  journal={PLoS One},
  volume={9},
  number={1},
  pages={e86746},
  year={2014},
  publisher={Public Library of Science San Francisco, USA}
}

@article{huang2017processes,
  title={Processes on the emergent landscapes of biochemical reaction networks and heterogeneous cell population dynamics: differentiation in living matters},
  author={Huang, Sui and Li, Fangting and Zhou, Joseph X and Qian, Hong},
  journal={Journal of the Royal Society Interface},
  volume={14},
  number={130},
  pages={20170097},
  year={2017},
  publisher={The Royal Society}
}

@article{morris2014mathematical,
  title={Mathematical approaches to modeling development and reprogramming},
  author={Morris, Rob and Sancho-Martinez, Ignacio and Sharpee, Tatyana O and Belmonte, Juan Carlos Izpisua},
  journal={Proceedings of the National Academy of Sciences},
  volume={111},
  number={14},
  pages={5076--5082},
  year={2014},
  publisher={National Acad Sciences}
}

@article{wang2016geometrical,
  title={A geometrical approach to control and controllability of nonlinear dynamical networks},
  author={Wang, Le-Zhi and Su, Ri-Qi and Huang, Zi-Gang and Wang, Xiao and Wang, Wen-Xu and Grebogi, Celso and Lai, Ying-Cheng},
  journal={Nature communications},
  volume={7},
  number={1},
  pages={1--11},
  year={2016},
  publisher={Nature Publishing Group}
}

@article{su2019phenotypic,
  title={Phenotypic heterogeneity and evolution of melanoma cells associated with targeted therapy resistance},
  author={Su, Yapeng and Bintz, Marcus and Yang, Yezi and Robert, Lidia and Ng, Alphonsus HC and Liu, Victoria and Ribas, Antoni and Heath, James R and Wei, Wei},
  journal={PLoS computational biology},
  volume={15},
  number={6},
  pages={e1007034},
  year={2019},
  publisher={Public Library of Science San Francisco, CA USA}
}

@article{angermueller2016deep,
  title={Deep learning for computational biology},
  author={Angermueller, Christof and P{\"a}rnamaa, Tanel and Parts, Leopold and Stegle, Oliver},
  journal={Molecular systems biology},
  volume={12},
  number={7},
  pages={878},
  year={2016}
}

@article{lecun2015deep,
  title={Deep learning},
  author={LeCun, Yann and Bengio, Yoshua and Hinton, Geoffrey},
  journal={nature},
  volume={521},
  number={7553},
  pages={436--444},
  year={2015},
  publisher={Nature Publishing Group}
}

@article{hornik1989multilayer,
  title={Multilayer feedforward networks are universal approximators.},
  author={Hornik, Kurt and Stinchcombe, Maxwell and White, Halbert and others},
  journal={Neural networks},
  volume={2},
  number={5},
  pages={359--366},
  year={1989}
}

@article{hinton2006reducing,
  title={Reducing the dimensionality of data with neural networks},
  author={Hinton, Geoffrey E and Salakhutdinov, Ruslan R},
  journal={science},
  volume={313},
  number={5786},
  pages={504--507},
  year={2006},
  publisher={American Association for the Advancement of Science}
}

@article{raissi2019physics,
  title={Physics-informed neural networks: A deep learning framework for solving forward and inverse problems involving nonlinear partial differential equations},
  author={Raissi, Maziar and Perdikaris, Paris and Karniadakis, George E},
  journal={Journal of Computational Physics},
  volume={378},
  pages={686--707},
  year={2019},
  publisher={Elsevier}
}

@article{chen2020solving,
  title={Solving inverse stochastic problems from discrete particle observations using the fokker-planck equation and physics-informed neural networks},
  author={Chen, Xiaoli and Yang, Liu and Duan, Jinqiao and Karniadakis, George Em},
  journal={arXiv preprint arXiv:2008.10653},
  year={2020}
}

@article{raissi2020hidden,
  title={Hidden fluid mechanics: Learning velocity and pressure fields from flow visualizations},
  author={Raissi, Maziar and Yazdani, Alireza and Karniadakis, George Em},
  journal={Science},
  volume={367},
  number={6481},
  pages={1026--1030},
  year={2020},
  publisher={American Association for the Advancement of Science}
}

@article{lecun1995convolutional,
  title={Convolutional networks for images, speech, and time series},
  author={LeCun, Yann and Bengio, Yoshua and others},
  journal={The handbook of brain theory and neural networks},
  volume={3361},
  number={10},
  pages={1995},
  year={1995}
}

@inproceedings{rahaman2019spectral,
  title={On the spectral bias of neural networks},
  author={Rahaman, Nasim and Baratin, Aristide and Arpit, Devansh and Draxler, Felix and Lin, Min and Hamprecht, Fred and Bengio, Yoshua and Courville, Aaron},
  booktitle={International Conference on Machine Learning},
  pages={5301--5310},
  year={2019},
  organization={PMLR}
}

@article{bonazzi2014symmetry,
  title={Symmetry breaking in spore germination relies on an interplay between polar cap stability and spore wall mechanics},
  author={Bonazzi, Daria and Julien, Jean-Daniel and Romao, Maryse and Seddiki, Rima and Piel, Matthieu and Boudaoud, Arezki and Minc, Nicolas},
  journal={Developmental cell},
  volume={28},
  number={5},
  pages={534--546},
  year={2014},
  publisher={Elsevier}
}

@incollection{davis2011remarks,
  title={Remarks on some nonparametric estimates of a density function},
  author={Davis, Richard A and Lii, Keh-Shin and Politis, Dimitris N},
  booktitle={Selected Works of Murray Rosenblatt},
  pages={95--100},
  year={2011},
  publisher={Springer}
}

@article{song2016accurate,
  title={Accurate cervical cell segmentation from overlapping clumps in pap smear images},
  author={Song, Youyi and Tan, Ee-Leng and Jiang, Xudong and Cheng, Jie-Zhi and Ni, Dong and Chen, Siping and Lei, Baiying and Wang, Tianfu},
  journal={IEEE transactions on medical imaging},
  volume={36},
  number={1},
  pages={288--300},
  year={2016},
  publisher={IEEE}
}

@article{limpert2001log,
  title={Log-normal distributions across the sciences: keys and clues},
  author={Limpert, Eckhard and Stahel, Werner A and Abbt, Markus},
  journal={BioScience},
  volume={51},
  number={5},
  pages={341--352},
  year={2001},
  publisher={American Institute of Biological Sciences}
}

@article{drake2013model,
  title={Model of fission yeast cell shape driven by membrane-bound growth factors and the cytoskeleton},
  author={Drake, Tyler and Vavylonis, Dimitrios},
  journal={PLoS Comput Biol},
  volume={9},
  number={10},
  pages={e1003287},
  year={2013},
  publisher={Public Library of Science}
}

@article{banerjee2016shape,
  title={Shape dynamics of growing cell walls},
  author={Banerjee, Shiladitya and Scherer, Norbert F and Dinner, Aaron R},
  journal={Soft matter},
  volume={12},
  number={14},
  pages={3442--3450},
  year={2016},
  publisher={Royal Society of Chemistry}
}

@article{udrescu2021symbolic,
  title={Symbolic pregression: Discovering physical laws from distorted video},
  author={Udrescu, Silviu-Marian and Tegmark, Max},
  journal={Physical Review E},
  volume={103},
  number={4},
  pages={043307},
  year={2021},
  publisher={APS}
}

@article{johnson2017building,
  title={Building a 3D integrated cell},
  author={Johnson, Gregory R and Donovan-Maiye, Rory M and Maleckar, Mary M},
  journal={bioRxiv},
  pages={238378},
  year={2017},
  publisher={Cold Spring Harbor Laboratory}
}

@article{campas2009shape,
  title={Shape and dynamics of tip-growing cells},
  author={Campas, Otger and Mahadevan, L},
  journal={Current Biology},
  volume={19},
  number={24},
  pages={2102--2107},
  year={2009},
  publisher={Elsevier}
}

@article{Cavanagh2021Physics,
  title={Physics-Informed Deep Learning Characterizes Morphodynamics of Asian Soybean Rust Disease, Morphodynamics},
  author={Cavanagh, Henry and Mosbach, Andreas and Scalliet, Gabriel and Lind, Rob and Endres, Robert G.},
  note={DOI:10.5281/zenodo.5525043},
  year={2021},
}

@article{kingma2014adam,
  title={Adam: A method for stochastic optimization},
  author={Kingma, Diederik P and Ba, Jimmy},
  journal={arXiv preprint arXiv:1412.6980},
  year={2014}
}

@article{kingma2013auto,
  title={Auto-encoding variational bayes},
  author={Kingma, Diederik P and Welling, Max},
  journal={arXiv preprint arXiv:1312.6114},
  year={2013}
}

@article{abcsmc,
  title={Approximate Bayesian computation scheme for parameter inference and model selection in dynamical systems},
  author={Toni, Tina and Welch, David and Strelkowa, Natalja and Ipsen, Andreas and Stumpf, Michael PH},
  journal={Journal of the Royal Society Interface},
  volume={6},
  number={31},
  pages={187--202},
  year={2009},
  publisher={The Royal Society London}
}

@article{takeshita2011role,
  title={On the role of microtubules, cell end markers, and septal microtubule organizing centres on site selection for polar growth in Aspergillus nidulans},
  author={Takeshita, Norio and Fischer, Reinhard},
  journal={Fungal biology},
  volume={115},
  number={6},
  pages={506--517},
  year={2011},
  publisher={Elsevier}
}

@article{van2008visualizing,
  title={Visualizing data using t-SNE.},
  author={Van der Maaten, Laurens and Hinton, Geoffrey},
  journal={Journal of machine learning research},
  volume={9},
  number={11},
  year={2008}
}

@article{xu2020solving,
  title={Solving Fokker-Planck equation using deep learning},
  author={Xu, Yong and Zhang, Hao and Li, Yongge and Zhou, Kuang and Liu, Qi and Kurths, J{\"u}rgen},
  journal={Chaos: An Interdisciplinary Journal of Nonlinear Science},
  volume={30},
  number={1},
  pages={013133},
  year={2020},
  publisher={AIP Publishing LLC}
}

@article{lu2021deepxde,
  title={DeepXDE: A deep learning library for solving differential equations},
  author={Lu, Lu and Meng, Xuhui and Mao, Zhiping and Karniadakis, George Em},
  journal={SIAM Review},
  volume={63},
  number={1},
  pages={208--228},
  year={2021},
  publisher={SIAM}
}

@article{klinger2018pyabc,
  title={pyABC: distributed, likelihood-free inference},
  author={Klinger, Emmanuel and Rickert, Dennis and Hasenauer, Jan},
  journal={Bioinformatics},
  volume={34},
  number={20},
  pages={3591--3593},
  year={2018},
  publisher={Oxford University Press}
}

@book{bradski2008learning,
  title={Learning OpenCV: Computer vision with the OpenCV library},
  author={Bradski, Gary and Kaehler, Adrian},
  year={2008},
  publisher={" O'Reilly Media, Inc."}
}

@inproceedings{abadi2016tensorflow,
  title={Tensorflow: A system for large-scale machine learning},
  author={Abadi, Mart{\'\i}n and Barham, Paul and Chen, Jianmin and Chen, Zhifeng and Davis, Andy and Dean, Jeffrey and Devin, Matthieu and Ghemawat, Sanjay and Irving, Geoffrey and Isard, Michael and others},
  booktitle={12th $\{$USENIX$\}$ symposium on operating systems design and implementation ($\{$OSDI$\}$ 16)},
  pages={265--283},
  year={2016}
}

\end{document}


\pagenumbering{gobble} 
\maketitle
\begin{center}
Henry Cavanagh, Andreas Mosbach, Gabriel Scalliet, Rob Lind, Robert G. Endres* \\
*Corresponding author: r.endres@imperial.ac.uk
\end{center}

\tableofcontents
\newpage

\pagenumbering{arabic}  

\section{Supplementary Figures}

\subsection{Supplementary Figure 1: Convergence of the physics-informed neural network (PINN)}

\begin{figure}[H]
\includegraphics[]{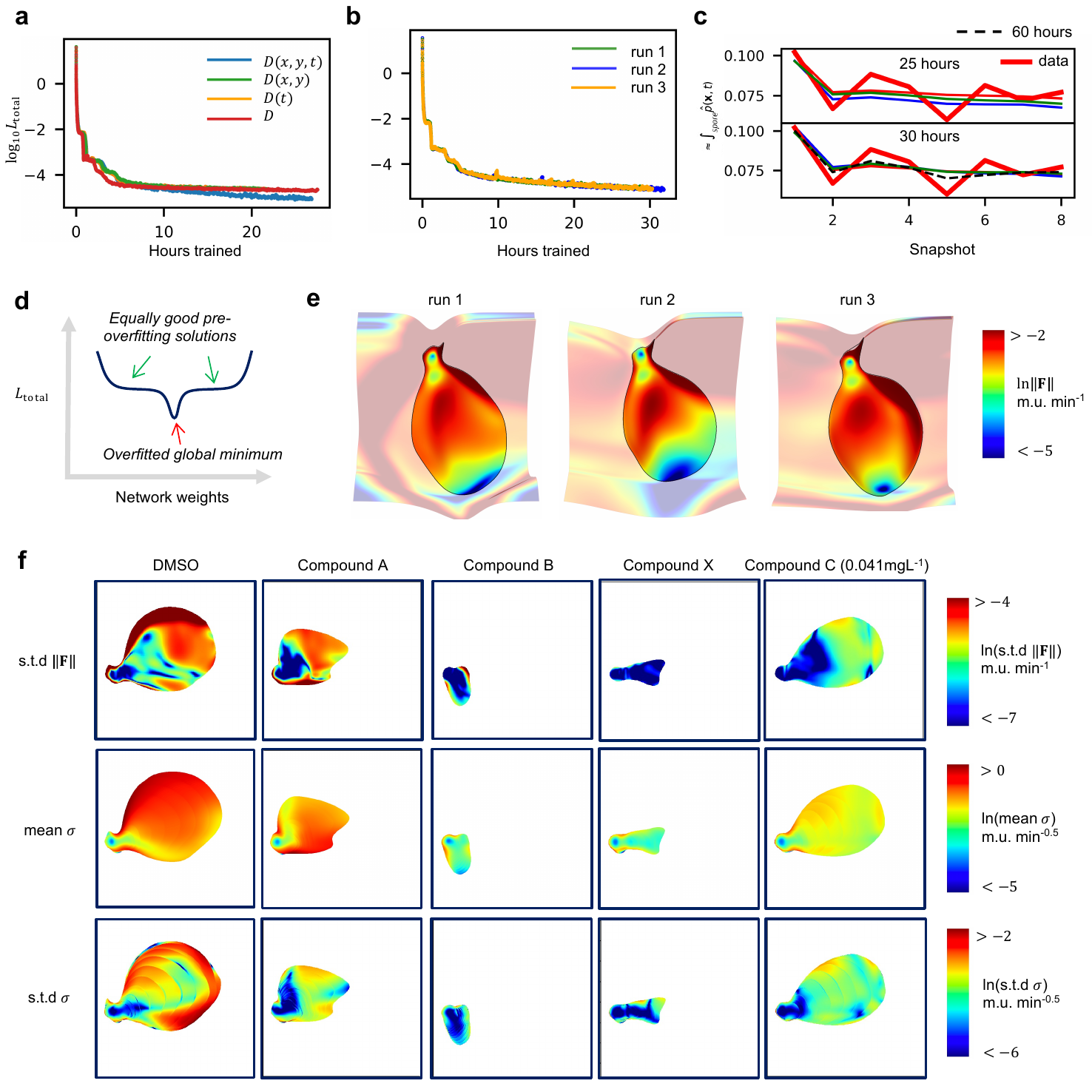}
\caption{\textbf{Convergence of the physics-informed neural network (PINN).} Caption continued on the following page.} 
\end{figure}

\begin{figure}[H]
  \contcaption{\textbf{Convergence of the physics-informed neural network (PINN).} (\textbf{a}) An ablation analysis comparing how $L\_{total}$ (running mean over 200 mini-batches) decreases as training progresses reveals that a diffusivity with both spatial and time dependence is the best model. The majority of the benefit likely comes from the dynamics in the spore region of morphospace, where diffusion is very high at first, and then strongly decreases such that not all spores germinate. (\textbf{b}) We repeat PINN network training three times for each condition, with different mini-batches, and $L\_{total}$ (running mean over 200 mini-batches) is shown here for each repeat of DMSO. (\textbf{c}) An approximation of the dynamics of the fraction of spores ($\approx \int\_{spore} \hat{p}(\mathbf{x}, t)$) across snapshots for DMSO (with the first snapshot not shown due to its much higher fraction of spores), found by numerically integrating a box around the spore PDF peak. The data is shown in red, and the three repeats have the same coloring as in (b). The PINN first explores smooth low-frequency solutions, fitting trends common to all snapshots, before ultimately beginning to overfit to the individual snapshots, as shown for one repeat at 60 hours in black. We stop training when the PINN begins to fit to the individual snapshots, which approximately corresponds to 30, 30, 30, 20, 25 and 25 hours for DMSO and Compounds A, B, C (0.041 mgL\textsuperscript{-1}), C (10 mgL\textsuperscript{-1}) and X, respectively. (\textbf{d}) Sketch of the loss landscape, whereby the global minimum is an overfitted solution, and there may be many equally good solutions before overfitting. (\textbf{e}) The landscapes from each of the three repeats after 30 hours of training for DMSO show many common features in the central data-rich region. (\textbf{f}) For each of the conditions with significant germination (i.e. excluding Compound C at 10 mgL\textsuperscript{-1}), three outputs are shown: the uncertainty in the force magnitude, $\| \mathbf{F}\|$, calculated from the standard deviation across the three training repeats; the mean $\sigma$ (from Eq. 1), averaged over time for the same training repeat as those of the landscapes shown in the other figures, and only calculated over regions where the PDFs are above $10^{-3}$; and the uncertainty in $\sigma$, calculated in the same way as the uncertainty for $\| \mathbf{F}\|$. All outputs are expressed in terms of morphospace units, m.u. Source data are provided for (\textbf{c}).}
\end{figure}


\subsection{Supplementary Figure 2: Correspondence between the landscapes and morphospace}

\begin{figure}[H]
\includegraphics[width=0.85\linewidth]{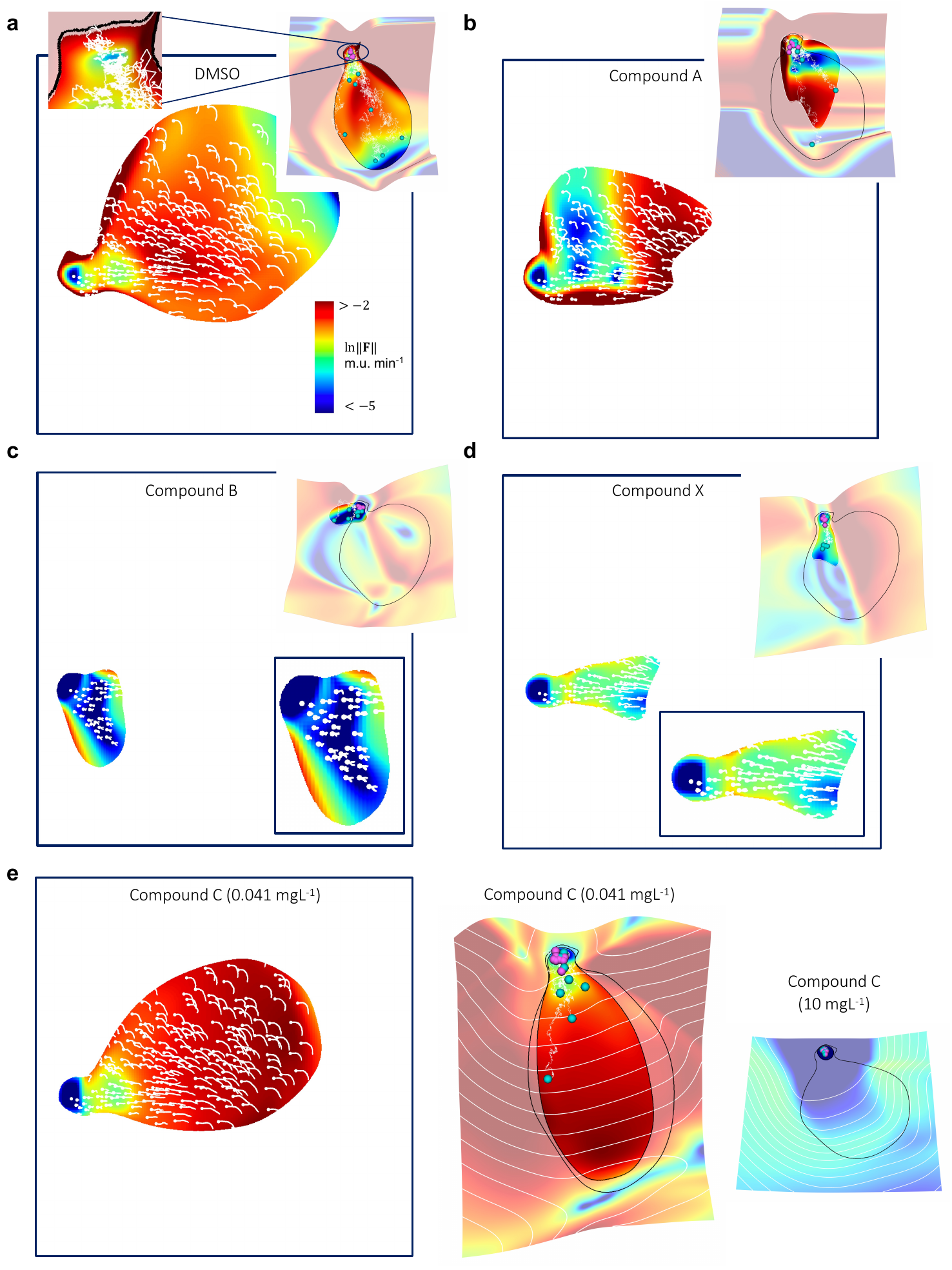}
\caption{\textbf{Correspondence between the landscapes and morphospace.} (\textbf{a-d}) Landscapes for DMSO and Compounds A, B and X, with samples of fungus images from the underlying morphospace, both colored by the gradient magnitude, $\| \mathbf{F}\|$, at regions where the PDFs are above $10^{-3}$. (\textbf{e}) The same as described above, but for Compound C at 0.041 mgL\textsuperscript{-1}, alongside the landscapes for Compound C at 0.041 mgL\textsuperscript{-1} (with contours along equal landscape values, spaced 0.14 m.u.\textsuperscript{2} min\textsuperscript{-1} apart, where m.u. stands for morphospace units) and 10 mgL\textsuperscript{-1} (with contours spaced 0.02 m.u.\textsuperscript{2} min\textsuperscript{-1} apart).}
\end{figure}

\subsection{Supplementary Figure 3: Correspondence between the landscapes and morphospace}

\begin{figure}[H]
\includegraphics[]{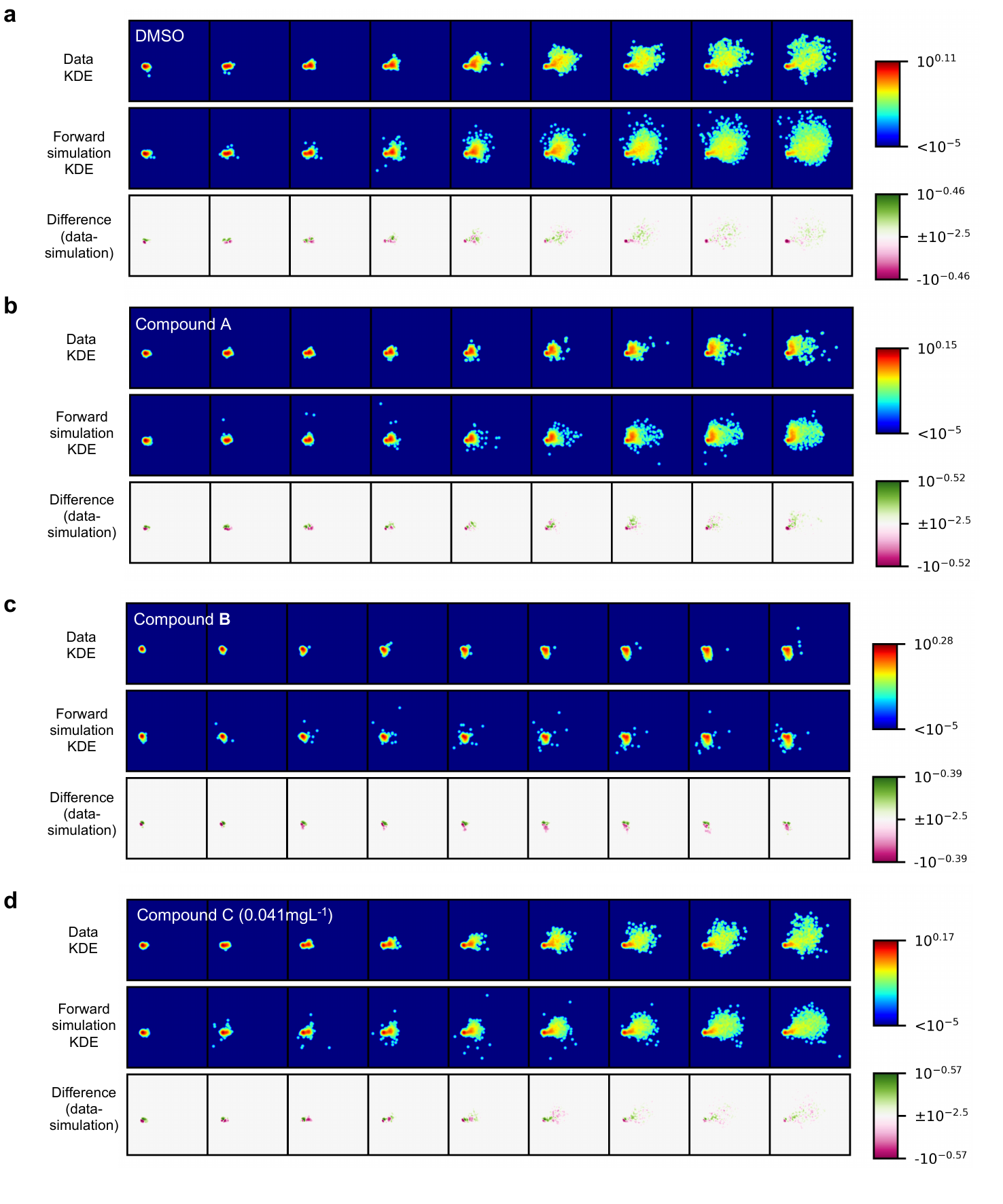}

\end{figure}

\begin{figure}[H]
\includegraphics[]{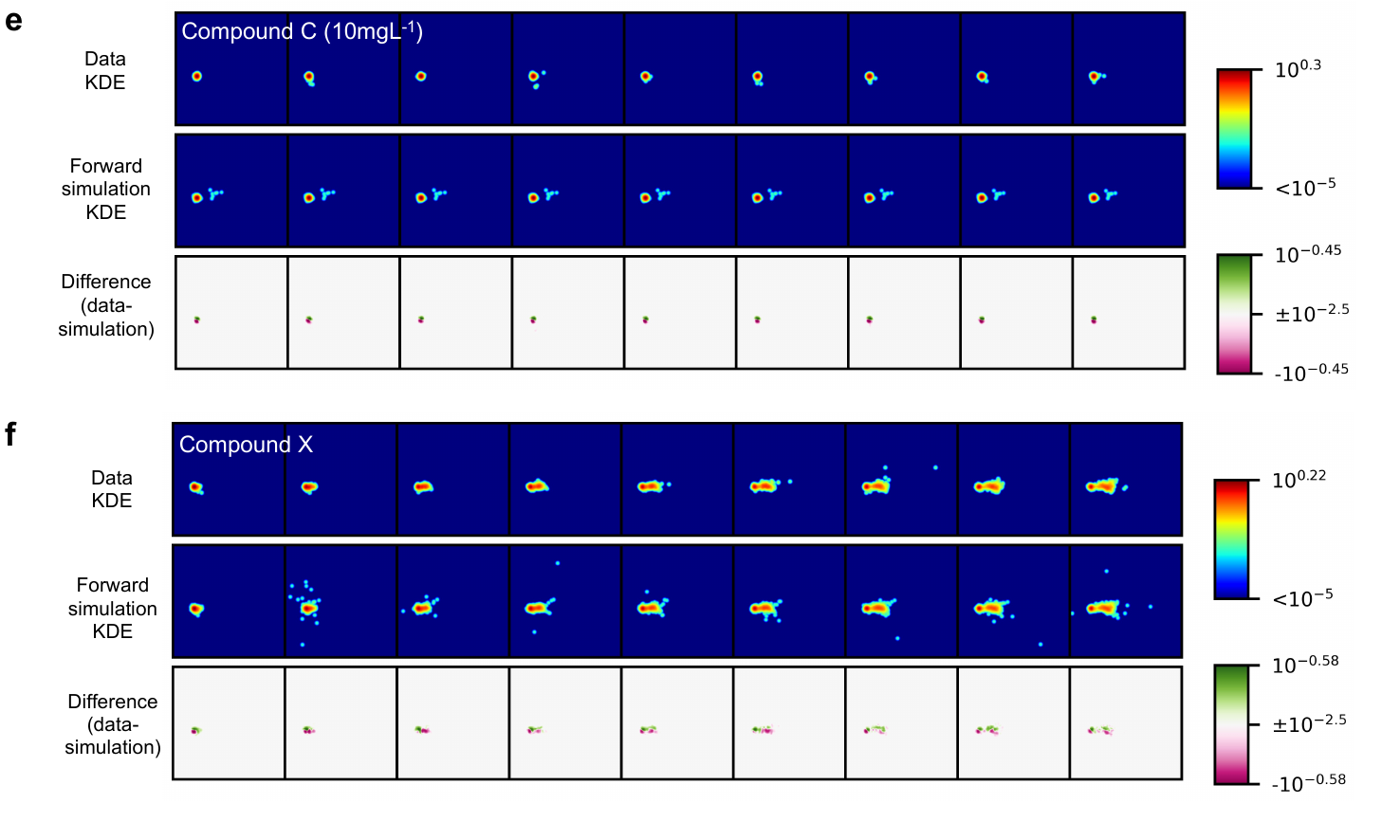}
\caption{\textbf{Validation of the landscapes and diffusivities learned by the PINN.} For each condition (\textbf{a}-\textbf{f}), three panels are shown: the first panel is the data kernel density estimate (KDE), the second is the KDE over simulations, and the third is the error (data KDE - simulation KDE). All are displayed on a logarithmic scale, and the error is truncated at $10^{-2.5}$, which is the probability density generated by a single particle, in order to highlight more systematic errors. For the forward simulations, particle starting positions were sampled from the initial probability distribution learned by the PINN, and then simulations were run by evaluating the potential and diffusivity on a $1000\times1000$ spatial grid, with 20 snapshots in time for the diffusivity. The figure shows good agreement across all conditions, validating the landscapes and diffusivities learned by the PINN.}
\end{figure}

\newpage

\subsection{Supplementary Figure 4: Comparison of simulation and data trajectories}

\begin{figure}[H]
\includegraphics[]{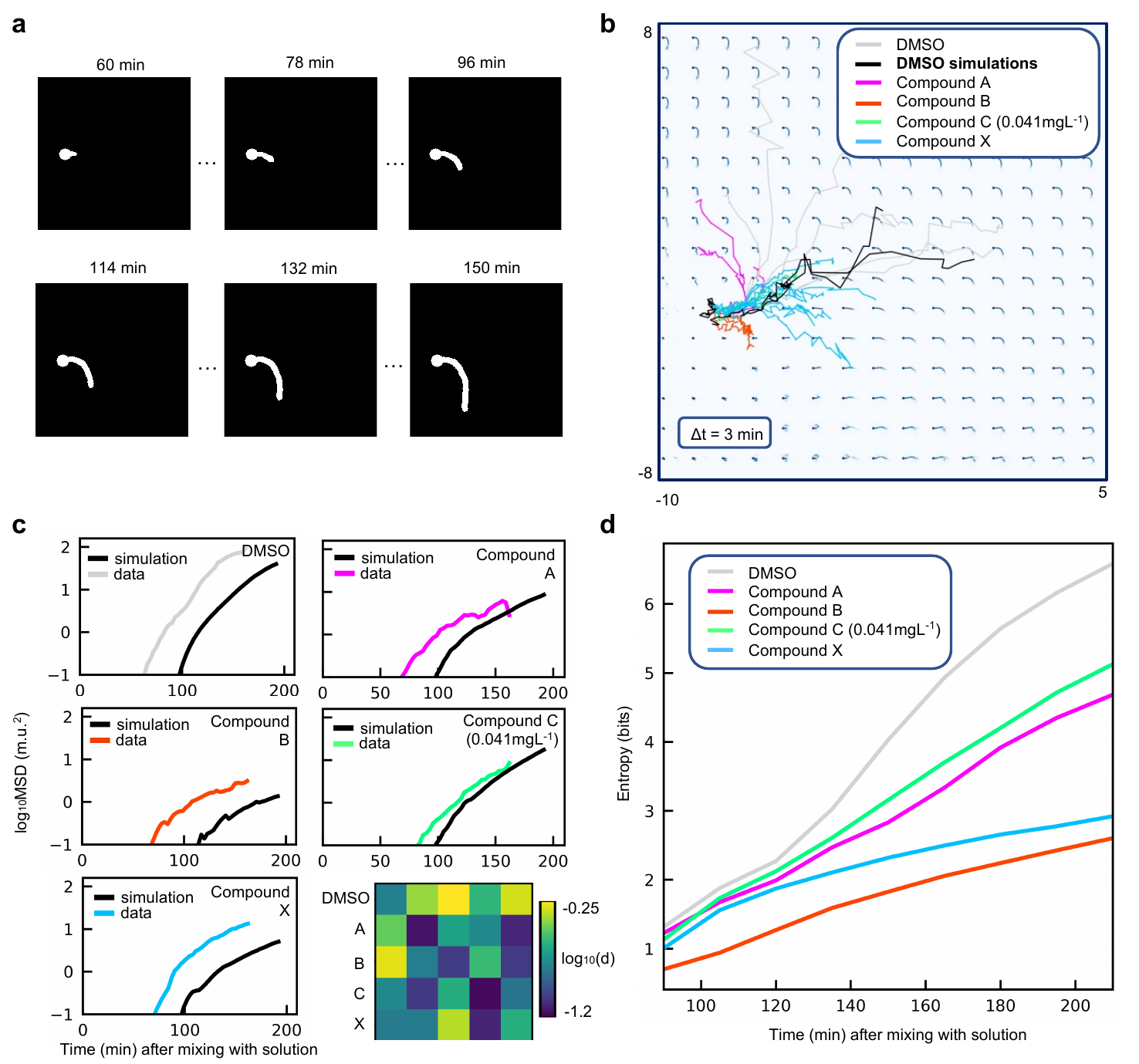}
\caption{\textbf{Comparison of simulation and data trajectories.}  (\textbf{a}) An example sequence from the DMSO time-lapse videos. Images were taken every 3 min, from 60 min after mixing with the compounds.  (\textbf{b}) Trajectories of sequential frames of the time-lapse videos (colored) and a sample of DMSO simulations (black). For the forward simulations, particle starting positions were sampled from the initial probability distribution learned by the PINN, and then simulations were run by evaluating the potential and diffusivity on a $1000\times1000$ spatial grid, with 20 snapshots in time for the diffusivity.  (\textbf{c}) Mean squared displacement (MSD, in terms of morphospace units, m.u.) plots against time for the time-lapse videos (colored) and forward simulations (black). Time-lapse videos were taken under higher temperatures, which results in early germination. A confusion matrix of the mean absolute differences ($d$) of the plots is also shown. For each simulation-data pairing (videos down the rows, simulations across columns), the time series were shifted horizontally and the result for each pairing taken to be the minimum of the mean absolute differences across the shifts. Simulations match their corresponding data generally the best, except for the simulations of Compound X. (\textbf{d}) The entropy of PDFs from a KDE over single particle simulations of Eq. 1 reveals that entropy always increases with time. Entropy is calculated as $-\sum_{x_{1}} \sum_{x_{2}} p(\mathbf{x}) \log_{2}p(\mathbf{x})\Delta x_{1} \Delta x_{2}$ with $\mathbf{x}=(x_{1}, x_{2})$ and only summing over morphospace regions where the PDFs are above $10^{-3}$. Source data are provided for (\textbf{c}-\textbf{d}).}
\end{figure}

\clearpage

\subsection{Supplementary Figure 5: Data-driven development of the tip growth model}

\begin{figure}[H]
\includegraphics[]{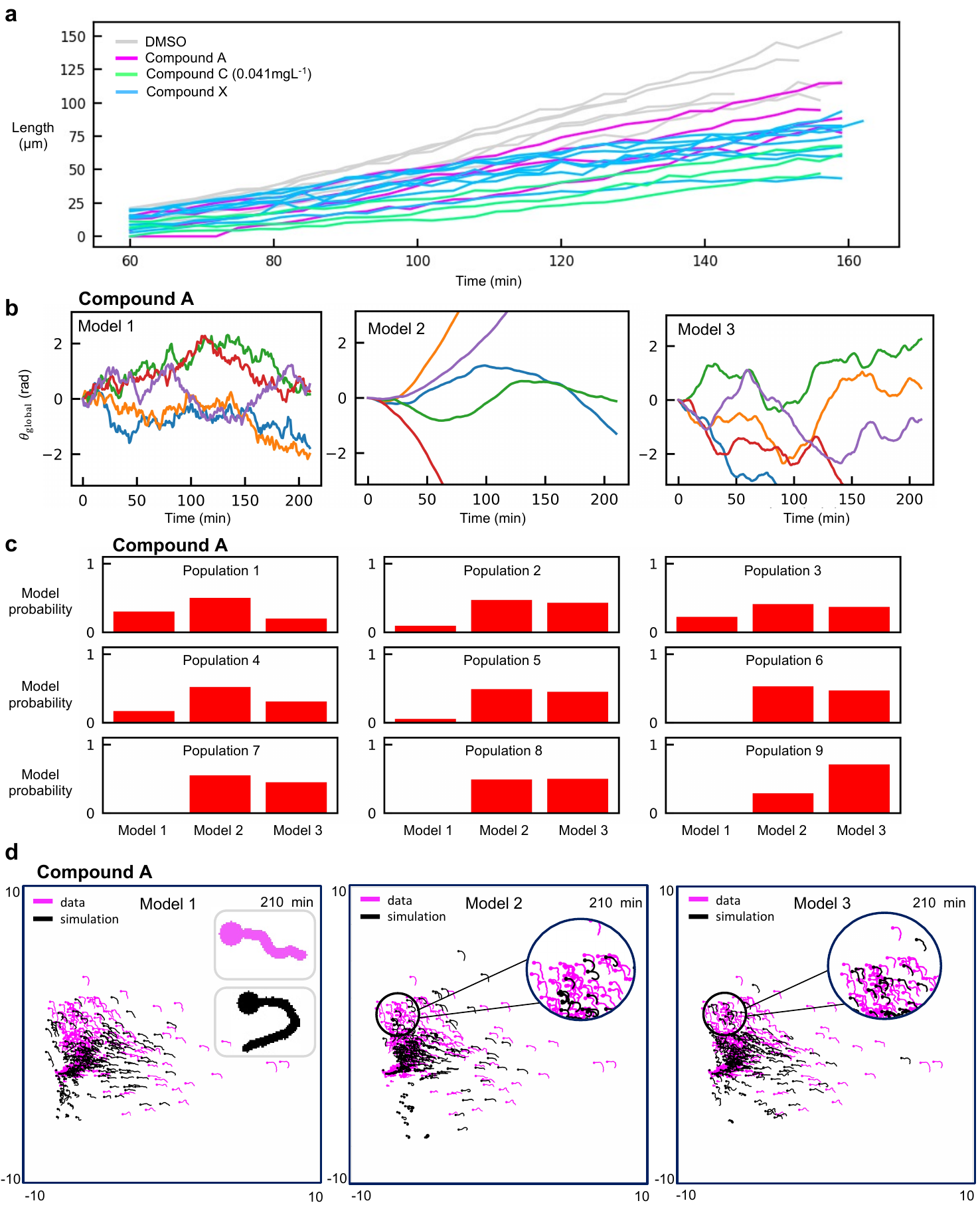}
\caption{\textbf{Data-driven development of the tip growth model.} Caption continued on the following page.}
\end{figure}

\begin{figure}[H]
  \contcaption{\textbf{Data-driven development of the tip growth model.} (\textbf{a}) Time-lapse video data shows length increasing approximately linearly with time for all conditions. (\textbf{b}) Variation in global direction, $\theta\_{global}$, for the three tip bending models tested (using bending MAP values for Compound A, with a growth rate of 0.75 $\mu$m min\textsuperscript{-1}). Model 1 is a random walk in $\theta\_{global}$, Model 2 is a random walk in path curvature, $\kappa$, and Model 3 is a persistent random walk in $\kappa$, with relaxation to straight growth. (\textbf{c}) Model selection using ABC-SMC. For early populations where the acceptance threshold is high, the lower dimensional parameter spaces of models 1 \& 2 lead to better fits. At lower acceptance thresholds, however, models 2 \& 3 fit better, validating model conception in the tip frame, and ultimately the relaxation to straight growth in Model 3 is required to reproduce the data distribution. (\textbf{d}) Comparison of Compound A snapshot data (pink) and MAP simulations (black) for the three models at 210 min, with an enlarged example of a randomly selected simulation and data fungus shown in the inset of the Model 1 box. While all models introduce bending too early for some fungi (the region below the spore where there are simulations but no data), Model 3 can reproduce the feature distribution best. In particular, it is the only model that can reproduce fungi with multiple bends in alternating directions. Overall, this feature is less well separated in this 2D morphospace that prioritizes global features. Source data are provided for (\textbf{a}, \textbf{c}).}
\end{figure}

\clearpage

\subsection{Supplementary Figure 6: Comparisons of maximum \textit{a posteriori} probability (MAP) simulations of Model 3 with data}

\begin{figure}[H]
\includegraphics[]{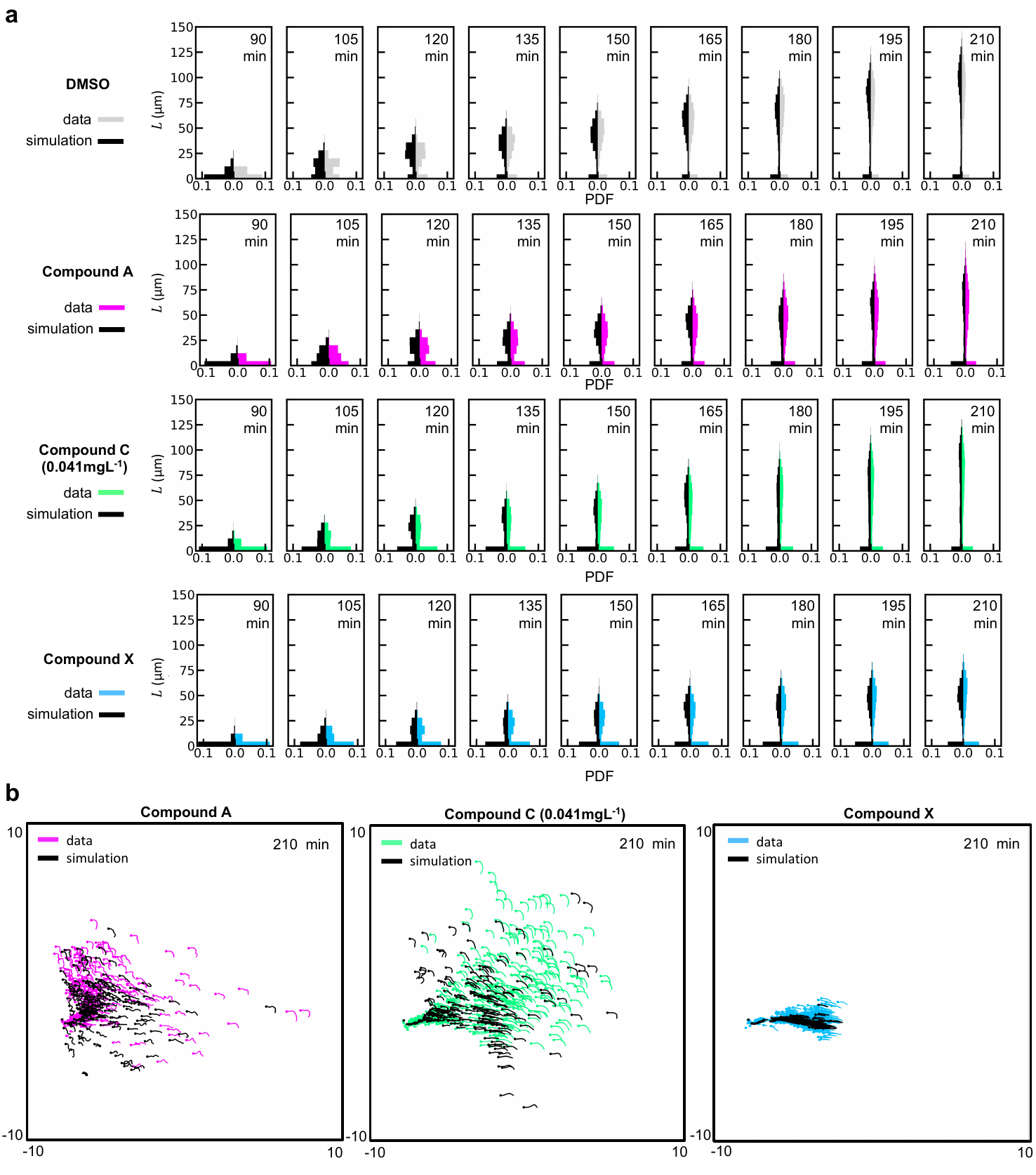}
\caption{\textbf{Comparisons of maximum \textit{a posteriori} probability (MAP) simulations of Model 3 with data.} (\textbf{a}) Comparisons of the length distributions for the snapshot data (colored) and simulations (black) with MAP parameters for all conditions, showing good agreement. (\textbf{b}) Comparisons of snapshot data (colored) and simulations (black) with MAP parameters at 210 min, showing good agreement for tip bending. Source data are provided for (\textbf{a}).}
\end{figure}

\clearpage

\subsection{Supplementary Figure 7: Chemical structure of Compound X}

\begin{figure}[H]
\includegraphics[]{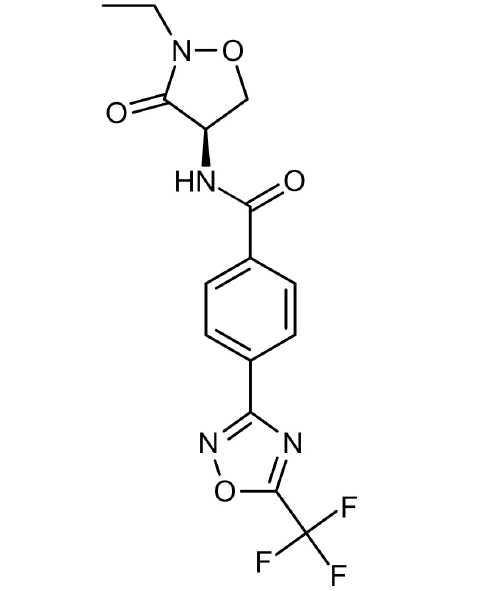}
\caption{\textbf{Chemical structure of Compound X.}}
\label{fig: FigureS7}
\end{figure}

\clearpage 


\section{Supplementary Notes}

\subsection{Supplementary Note 1: \textit{P. pachyrhizi} Imaging and Image Processing}

\textbf{\textit{P. pachyrhizi} spore propagation}

\vspace{2mm}

\noindent Urediniospores of the plant pathogenic basidiomycete \textit{P. pachyrhizi} Syd. \& P. Syd were prepared in a biosafety class 2 laboratory at Syngenta Stein, Switzerland, as follows: Glycine max (var. Toliman) plants were grown for 22-25 days in 8 cm pots in substrate with fertilizer. On the day before infection, all leaves except the second trifolium were cut, and plants were watered well. Approx. 30 mg of freshly harvested urediniospores of strain K8108 were suspended in 5 mL of 0.05 \% (v/v) Tween 20 solution in sterile water, and the suspension was diluted to a final concentration of 5$\times$10\textsuperscript{4} spores per mL in 100 mL Tween solution. The plants were inoculated inside a sterile bench by spraying the bottom sides of the leaves evenly using an airbrush. Infected plants were then incubated at 22 °C, \textgreater 90 \% relative humidity, in the dark. 24 h post infection, the plants were placed back under light (20 K Lux, Metal Halid) at 22 °C day / 20 °C night conditions, 70 \% relative humidity, and watered every 3 d. Once pustules (uredinia) were formed (usually 10 days post infection) the leaves were cut off and placed inside large Petri dishes containing a moisturized paper towel. The Petri dishes were closed with Parafilm and incubated for 3 to 4 d in the dark at room temperature. The spores were then detached from the upside-down held leaves inside a biosafety class 2 cabinet by gentle tapping, collected on sterile aluminum foil and used for microscopy assays on the same day.

\vspace{5mm}

\noindent \textbf{\textit{P. pachyrhizi in vitro} germination assay for snapshot and time-lapse imaging}

\vspace{2mm}

\noindent Chemicals tested were either provided by Syngenta or purchased from Merck KGaA (Darmstadt, Germany) or other vendors as indicated. Stock solutions were prepared at 10 gL\textsuperscript{-1} in DMSO (Fisher Chemical, D/4121/PB15): Carbendazim (CAS-number 10605-21-7; Merck, 45368-250MG), PIK-75 (2-Methyl-5-nitrobenzenesulfonic acid [(6-bromoimidazo[1,2-a]pyridin-3-yl)methylene]methylhydrazide hydrochloride; CAS-number 372196-77-5; Angene AG00C6HA / AGN-PC-0RDGQ1; stock solution stored frozen at -80 °C in small aliquots), Compound X (Syngenta research compound related to trifluoromethyloxadiazoles \cite{winter2020trifluoromethyloxadiazoles}), Benzovindiflupyr (CAS-number 1072957-71-1; Syngenta). All compounds tested were diluted to two-fold the final concentration in water, either directly from the concentrated stock solutions or from serial 3-fold dilutions in DMSO. Assay plates in 96-well format were prepared before the spore suspensions by pipetting 50 µL of two-fold concentrated treatment solution into the respective wells of CellCarrier-96 Ultra plates (PerkinElmer, 6055302).
Approx. 20 mg of freshly harvested \textit{P. pachyrhizi} urediniospores were suspended in 2 mL of 0.015 \% (v/v) Tween 20 in sterile water by shaking in a 5 mL vial (Axygen, SCT-5ML-S). 

After determining the spore concentration by using a hemocytometer (Neubauer improved), the required volume of two-fold concentrated suspension with 20,000 spores per mL in 0.0015 \% Tween 20 was prepared. The assay was started immediately by mixing 50 µL of spore suspension with treatment solution in the 96-well plates prepared in advance, resulting in a final one-fold concentration of the test compounds (Carbendazim: 1.1 mgL\textsuperscript{-1}; PIK-75: 3.3 mgL\textsuperscript{-1}; Compound X: 1.1 mgL\textsuperscript{-1}; Benzovindiflupyr: 0.041 and 10 mgL\textsuperscript{-1}) and 1000 spores per well, in a final DMSO concentration of 0.1 \% in all wells. The plates were then incubated in a closed box at 20 °C in the dark.
After 9 different incubation times between 90 and 210 min in 15 min intervals, 25 µL Calcofluor White staining solution (Merck, 18909-100ML-F) and 25 µL KOH 10 \% (w/v) were added to each well to stain the fungal cell walls and simultaneously kill the germlings to enable snapshot imaging. The plates were sealed with adhesive aluminium PCR sealing foil (Thermo Scientific, AB-0626), mixed well by vortexing, and left at least for 20 min at room temperature to ensure consistent staining and to allow the spores to settle.

Imaging on the Opera QEHS High-Content Screening System (PerkinElmer) was performed with the following settings for excitation of Calcofluor White fluorescence (cell walls of germ tubes): 405 nm laser (350 µW), 10x air objective lens, 40 ms exposure time, camera bandpass filter 450/50 nm, pixel binning 2. For the excitation of autofluorescence of spore hulls (to differentiate them from germ tubes and appressoria) the following settings were used: 488 nm laser (15900 µW), 10x air objective lens, 1,600 ms exposure time, camera bandpass filter 565/40 nm, pixel binning 2. Twelve technical replicate wells were imaged per condition, and 47 images were taken per well at different positions.

Germination time-lapse imaging was performed on the JuLI Stage Real-Time Cell History Recorder (NanoEnTek Inc.). The germination assay was set up as described above, but after mixing spores with treatment solutions the assay plates were incubated directly in the imaging device, which itself was placed inside of a climate cabinet set to 20 °C. Imaging started with an autofocus run, followed by 321 runs with 3 min intervals to cover germination between +1 and +17 h after setup. No replicate wells were prepared to minimize interval times, but 5 different spots were imaged per well. The transmission light LED and the 10x objective were used with settings: 15 ms exposure time, LED power “3”, brightness correction “10”. 

\vspace{5mm}

\noindent \textbf{Image Processing}

\vspace{2mm}

\noindent The size of the snapshot image sets necessitated fully-automated processing. We extracted fungus contours using adaptive binarization, where the threshold value varies based on the statistics of a surrounding window. We used $100\times100$ windows (the approximate size of lighting defects) out of full images of $503\times685$ pixels, and the threshold point was found from the cross correlation of this area with a Gaussian window, and shifted with biases of 30 and 0 out of the [0, 255] pixel range for the germ tube and spore images, respectively (found through trial and error). See the \textit{adaptiveThreshold} function in OpenCV. This yielded two images per view: one with spore contours, and another with germ tube contours. Adding these together then gave an image with full fungi. Contours in this combined image with an area above a threshold found by trial and error were removed as obvious overlapping fungi, and the remainder were cropped by finding the minimum bounding rectangle (using OpenCV's \textit{minAreaRect} function), and rotating to align with the pixel grid (using OpenCV's \textit{warpAffine} function with bilinear interpolation) with padding so all were $200\times200$ regions of interest (ROIs), to fit the largest fungi in the set. Incomplete fungi were also removed by detecting if any of their outline points touched the image border.

We then used a supervised convolutional neural network to remove overlapping morphologies, given a set of hand-labelled examples. Single-fungus images were then translated and rotated so the initial growth directions coincided, and a flip was executed if the right-most point of the fungus was higher than the germination point. To do this alignment, we utilized the fact that there was one image of the spores, and another of the germ tubes for each view, and that these two contours overlapped slightly. The fungi were all translated so the points at which the germ tube hit the spore coincided. This point was found to be the center of mass of the overlap of the spore and germ tube contours. Since the germ tubes grow out perpendicular to the spore surface, alignment of the initial germination direction was then achieved by rotating the line joining the two points where the germ tube contour hit the spore to be vertical. These two points were found by tracing from the point on the germ tube contour furthest from the spore - germ tube intersection point in opposite directions until the spore was hit. We replaced all spores with identical circles, so as to prioritize modeling of the germ tube; the resulting morphospace point is then widened into a spore region through the kernel density estimation.

We removed overlapping fungi, which we note means larger morphologies are more likely to be removed than smaller ones (e.g. ungerminated spores), leading to a slight bias towards smaller morphologies. However, the DMSO data shows that there is very little germination after the first snapshot, which would lead to an artificially increasing percentage of spores if the bias was significant, which Supplemetary Fig. 1c does not show; instead, the inter-snapshot variability is the much more dominant factor. In other words, the variance in the latent variables that cause inter-snapshot variability dominates over the slight bias induced by removing overlapping fungi. In future, however, this slight bias can be resolved by using more sophisticated segmentation algorithms that can resolve overlapping morphologies \cite{song2016accurate}, which would also increase the amount of data extracted per image.

For the time-lapse videos, we again binarized (using trial and error for a suitable threshold value, and without adaptive binarization, since there were not significant lighting defects), and found the fungus contours. We then found series of contours across frames whose center of mass were closest, and manually looked through these to find those that corresponded to tracking an individual fungus. We then used ImageJ to color over the spore with the background color (black background, white fungi), and aligned all germination points (found simply as the intersection point on the initial full-fungus contour image and the new contour image with the germ tube only). We then manually rotated and (where neccessary) flipped the images using Gimp, to ensure the initial growth directions coincided. Before being inputted into the autoencoder, snapshot and time-lapse video pixels were assigned to a value in the set \{0, 1\}.

\subsection{Supplementary Note 2: Autoencoder and PINN Neural Networks}

\textbf{Autoencoder architecture and training}

\vspace{2mm}

\noindent For the autoencoder's encoder, we used four convolutional layers with 16, 32, 64 and 16 feature maps, all with 3$\times$3 kernels, ReLU activations, batch normalization, and alternating stride sizes of 1 and 2. The decoder's structure  mirrored the encoder's, but with transposed convolutions. We used a sigmoid output activation and binary cross entropy loss, over mini-batches of 50 images, and trained for 4 epochs using the Adam optimizer \cite{kingma2014adam} with a learning rate of $10^{-4}$, which took 2 hours with a Quadro RTX 6000 GPU card. Training was stopped at the point at which the trajectories of the single-fungus videos were least complex. 

Principal Component Analysis (PCA) is a commonly-used linear dimensionality technique that finds a set of orthogonal features based on the data variance in different directions. While this method can find very interpretable features, non-linear methods can be more expressive, capturing curved data manifolds to describe the data in fewer dimensions. This non-linearity is particularly useful for our work, as it allows us to find a 2D morphospace, which then enables intuitive visualization of the Fokker-Planck model landscapes.

t-distributed stochastic neighbor embedding (t-SNE, \cite{van2008visualizing}) is another commonly-used non-linear dimensionality reduction algorithm that could be used for finding the morphospace, and for characterizing morphodynamics with the Fokker-Planck model. However, this algorithm is non-parametric, meaning that when new data is added, it must be re-run in full, and so is not well-suited for the implementation of approximate Bayesian inference on mechanistic model parameters.

If the autoencoder is too small (i.e. shallow or narrow), the embedding function may be too simple to capture curved data manifolds. For example, autoencoders with single hidden layer and linear activation functions can only learn linear features. However, if the network is too large, it can reconstruct the data without needing to closely describe the underlying manifold. This can be thought of as a form of overfitting. For higher-dimensional latent spaces, this can be a more significant problem, and so autoencoder variants that regularize the latent space are typically used, for example the Variational Autoencoder (VAE) \cite{kingma2013auto}, which imposes a Gaussian prior distribution over the latent space. These ensure the latent space is smooth, with meaningful features throughout. We found that the autoencoder architecture used here was suitable for capturing the data manifold, and avoided variants like the VAE because they warped fungus trajectories that were more natural with a standard autoencoder.

\vspace{5mm}

\noindent \textbf{Learning landscapes from snapshot data}

\vspace{2mm}

\noindent For the PINN, the loss function to be minimized comprizes four terms, with the first three calculated over random mini-batches of $N$ data points, and the final one over the full spatial grid of $M$ data points. The first is the mean squared difference between the learned PDF, $\hat{p}(\mathbf{x}^{j},t^{j})$, and data, $p(\mathbf{x}^{j},t^{j})$,
\begin{equation}
    L\_{PDF} = \frac{1}{N} \sum_{j=1}^{N}{[ \hat{p}(\mathbf{x}^{j},t^{j}) - p(\mathbf{x}^{j},t^{j}) ]}^{2},
\end{equation}
with $\{\mathbf{x}^{j},t^{j}\}$ in the nine snapshots. The second is the mean squared PDF at the boundary, 
\begin{equation}
    L\_{BC} = \frac{1}{N} \sum_{j=1}^{N}{[\hat{p}(\mathbf{x}^{j},t^{j}) ]}^{2}
\end{equation}
with $\{\mathbf{x}^{j},t^{j}\}$ selected from $10^{6}$ uniformly distributed boundary points, and the third term is the mean squared PDE residual ($\mathcal{N}$, given in Eq. 3),
\begin{equation}
    L\_{PDE} = \frac{1}{N} \sum_{j=1}^{N}{[ \mathcal{N}(\hat{p}(\mathbf{x}^{j},t^{j}), \hat{D}(\mathbf{x}^{j},t^{j}) , \hat{U}(\mathbf{x}^{j})) ]}^{2},
\end{equation}
with $\{\mathbf{x}^{j},t^{j}\}$ selected from $10^{6}$ points uniformly distributed over the whole domain. The final term ensures the PDF integrates to one:
\begin{equation}
    L\_{norm} =  \left[ \sum_{j=1}^{M}{\Delta x_{1}\Delta x_{2} \hat{p}(\mathbf{x}^{j},t)} -1  \right]^{2},
\end{equation}
with $\mathbf{x}^{j}$ covering the full spatial grid and $t$ randomly selected.

For the total loss (Eq. 4), we used hyperparameters of 1, 1, 500, 0.01 for $a$, $b$, $c$ and $d$. Previous work using PINNs to solve the Fokker-Planck equation \cite{xu2020solving, chen2020solving} used a weighting for $L\_{PDE}$ of 100, and the same values for the other hyperparameters as we used. Since our data was of snapshots of different spore batches, we increased the weighting to 500, to prioritize more the PDE fitting over the data. While initially selected by trial and error through inspecting single particle forward simulations, this choice was later vindicated quantitatively. We trained the DMSO PINN with $c$ values of 50, 500 and 5000 for 10 hours each. Since changing the hyperparameters modifies the loss calculation, $L\_{total}$ cannot be used to test the quality of the inferred parameters. Instead, we compared the resulting fits of single particle simulations of Eq. 1 with the data through the mean absolute probability error on the grid (MAE). Particle starting positions were sampled from the initial PDF learned by the PINN, and then simulations were run by evaluating the potential and diffusivity on a $1000\times1000$ spatial grid, with 20 snapshots in time for the diffusivity, with a time step of 0.01 min. The MAE values for $c$ values of 50, 500 and 5000 were $9.8\times10^{-6}$, $9.0\times10^{-6}$ and $9.9\times10^{-6}$ respectively, showing 500 was a suitable choice for $c$, as it produced the model with the best fit to the data. Reducing the relative weighting of the data penalty means there is no longer a guarantee of the learned PDF integrating to one, and so a normalization penalty is required. The weighting for $L\_{norm}$ of 0.01 was found to normalize the data adequately, without excessively prioritising this constraint, in agreement with the findings of \cite{xu2020solving, chen2020solving}. 

Since \textit{P. pachyrhizi} were fixated upon staining, the snapshots were of distinct populations. There are therefore latent variables that change the dynamics for each population. Ideally, the learned solution would be the average dynamics of infinite repeats. Since neural networks learn smooth low-frequency solutions first, we used early stopping to capture average dynamics, rather than letting the dynamics fit exactly to each snapshot, primarily watching for when the PINN began to fit individual snapshots. This point approximately corresponded to 30, 30, 30, 20, 25 and 25 hours for DMSO and Compounds A, B, C (0.041 mgL\textsuperscript{-1}), C (10 mgL\textsuperscript{-1}) and X, respectively. For small numbers of snapshots, it is possible that spurious patterns emerge not only on the single-snapshot level, but also across many snapshots; for example if two subsequent snapshots have similar low-probability latent variables, this will be captured by the PINN. However, such patterns become increasingly less problematic as the number of snapshots increases. An alternative solution is to constrain the solution physically, as we did with the tip growth model.

The three neural networks had 5 fully connected layers, each with 50 neurons, with residual skip connections, and swish activations between layers. A softplus output activation was used for the PDF, and sigmoid was used for the potential and diffusivity, with the potential multiplied by 3 to give a [0, 3] output range. This sigmoid constraint prevents unphysical solutions that can arise with unbounded force and diffusivity. Output variables that share inputs (e.g. the PDF and diffusivity) can be outputted from a single neural network if they are likely to have similar features, for increased computational efficiency. We used the Adam optimizer \cite{kingma2014adam} with a learning rate of $5\times10^{-4}$, and batch sizes, $N$, of 8,000. To speed up training, the DMSO landscape was first trained for 10 hours, and PINNs for the other conditions were initialized with these weights (known as transfer learning).

The landscapes for DMSO and Compound A formed deep bowls at the outer morphospace regions. For visualization, we therefore scaled landscape regions outside the PDF boundary of $10^{-3}$ towards the closest point on this boundary and applied a Gaussian smoothing to this outer region. This way, the full landscape can be easily compared with the morphospace, but with decreased gradients in regions where there is not significant data density. This is a purely cosmetic choice, and the original landscapes can also be viewed unedited, since the translucency does enable viewing of valley regions.

There are two sources of stochasticity in the fungus trajectories over the landscapes, which we call morphodynamic diffusion and embedding noise, both of which are consumed into the diffusion term of the Fokker-Planck model. Morphodynamic diffusion is the fundamental unpredictability of morphodynamics over time, arising from un-modeled internal and external factors that vary across fungi. This corresponds to diffusion over the approximately 2D morphological manifold within the high-dimensional pixel space. Embedding noise arises from fungi being randomly perturbed away from the 2D morphology manifold in the high-dimensional pixel space. Such perturbations arise from factors including image resolution, segmentation and alignment (which for example add extra degrees of freedom to the images, and each degree of freedom expand the data manifold dimensionality), and the complexity of the autoencoder's embedding function means these perturbations are not always just mapped to the closest point on the manifold. Since the same image pre-processing algorithms were used on all conditions, differences in the Fokker-Planck diffusion over the same region of morphospace will be morphodynamic in nature, rather than arising from the embedding noise.

\subsection{Supplementary Note 3: Three Models of Tip Growth}

The minimal model for fungus growth was composed of two equations: one for lengthening and another for tip bending, and we ran model selection on three candidate models for the bending part.

For fungus lengthening, we used the 3-parameter lognormal distribution to model both germination time, $t_g$, and growth rate, $\alpha$. The probability density function of the 3-parameter lognormal distribution is given by
\begin{equation}
    f(x; s, \sigma^{2}, \text{loc}) = \frac{1}{\sigma\sqrt{2\pi}(x-\text{loc})}\exp{\frac{\log^{2}\left(\frac{x-\text{loc}}{s}\right)}{2\sigma^2}}
\end{equation}
where $\sigma$ is a shape parameter, $s$ is a scale parameter (also the median), and $\text{loc}$ is a location parameter (the lower bound). The 2-parameter distribution has $\text{loc}$ set to zero.

We modeled germination time, $t_g$, as distributed according to $t\_{g} \sim lognormal(s_{t\_{g}}, \sigma_{t\_{g}}, \text{loc}_{t\_{g}})$, and growth rate, $\alpha$, as distributed according to $\alpha = \text{loc}_{\alpha}-x$, with $x \sim lognormal(s_{\alpha}, \sigma_{\alpha}, 0)$ and resampling for negative $\alpha$.

For inferring both the lengthening and bending parameters, we used ABC-SMC \cite{abcsmc}. This is a computationally efficient implementation of ABC, identifying intermediate distributions over a series of populations, and gradually decreasing the acceptance threshold. All histograms were compared using the summed absolute distance, and we trained the autoencoder for an extra two epochs with simulations generated randomly from the prior distribution to get coverage of any novel features.

For inferring lengthening parameters ($s_{t\_{g}}$, $\sigma_{t\_{g}}$, $\text{loc}_{t\_{g}}$, $s_{\alpha}$, $\sigma_{\alpha}$, $\text{loc}_{\alpha}$), we ran ABC-SMC with a population size of 100, each with 5000 simulations, for 4 steps. All prior distributions were uniform distributions over the following ranges: $s_{t\_{g}}$: [0.01, 150], $\sigma_{t\_{g}}$: [0.01, 5], $\text{loc}_{t\_{g}}$: [20, 100], $s_{\alpha}$: [0.01, 2], $\sigma_{\alpha}$: [0.01, 1], $\text{loc}_{\alpha}$: [0.3, 1.5].

We compared three models for tip bending, and used ABC-SMC for both model selection and parameter inference. As described in the main text, the comparison was done using both the fungus length data and 210 min morphospace embeddings, and Model 3 was found to reproduce the data best. For all of the following, $\sigma$ is a noise parameter that was fitted, and $dW$ is the Wiener process. Model 1 was a random walk in the global direction, $\theta\_{global}$, a simple model commonly used in the literature:
\begin{equation}
    d\theta\_{global} = \sigma dW.
\end{equation}
Model 2 was a random walk in the curvature of the growth path, $\kappa$, in order to connect to cell tip mechanics:
\begin{equation}
    d\kappa = \sigma dW.
\end{equation}
Model 3 was a persistent random walk in the curvature, with an additional parameter for relaxation to straight growth, $\tau^{-1}$, motivated by work analysing fission yeast tip growth mechanics \cite{drake2013model}:
\begin{equation}
    d\kappa = -\tau^{-1}\kappa dt + \sigma dW.
\end{equation}

Models 2 and 3 can be loosely connected to a diffusing growth zone at the tip as has been described in fission yeast \cite{drake2013model}, by introducing an angular growth zone position, $\theta\_{tip}$, and a mapping, $\kappa = f(\theta\_{tip})$, where $f$ is unknown, but likely monotonically increasing, and passing through the origin (i.e. a central growth zone corresponds to straight growth).

For fitting the bending parameters using ABC-SMC, images were created by taking the MAP lengthening parameters, running the model, and then converting to Cartesian coordinates using the relation $\frac{d\theta\_{global}}{dt}=\frac{dL}{dt}\frac{d\theta\_{global}}{dL}=\alpha\kappa$, such that: 
\begin{equation}
    x_{N+1} = x_{N} + \alpha\Delta t \cos\left(\textstyle \sum_{n=0}^{N}\alpha\kappa\Delta t\right)
\end{equation}
\begin{equation}
    y_{N+1} = y_{N} + \alpha\Delta t \sin\left(\textstyle \sum_{n=0}^{N}\alpha\kappa\Delta t\right).
\end{equation}
Coordinates were then converted to images using the \textit{polylines} function in OpenCV.

For selecting the optimal bending model, we ran ABC-SMC with a population size of 40, each with 1000 simulations, for 9 steps. All prior distributions were uniform distributions over the following ranges: Model 1, $\sigma$: [0, 0.2]; Model 2, $\sigma$: [0, 0.01]; Model 3, $\sigma$: [0, 0.05], $\tau^{-1}$: [0, 0.2]. 

For subsequently inferring the bending parameters of Model 3, we used a population size of 40, each with 1000 simulations, and ran for 4 steps using the same prior distributions as for model selection.

\subsection{Supplementary Note 4: Possible Modes of Action}

The compounds were identified at pre-screening by eye to show a range of phenotypes.

Compound A (methyl benzimidazol-2-ylcarbamate) is a widely-used fungicide that inhibits the assembly of tubulin subunits into functional microtubules, which are an essential part of the cytoskeleton \cite{takeshita2011role}. Microtubules participate in maintaining the shape of cells, the distribution of organelles, the transport of materials within the cell, and in the separation of chromosomes during mitosis. The bending of germ tubes that we observe may be in part induced by disturbed vesicle transport towards the growing hyphal tip, although there is likely a range of causes.

Compound B (PIK-75 hydrochloride) is a phosphoinositide 3-kinase (PI3K) inhibitor. PI3Ks are components of certain signaling pathways, which control growth, metabolism and other functions. PIK-75 comes from the pharmaceutical industry and is technically not a fungicide. The fungal target is unknown, but if it inhibits fungal PI3K, then the observed phenotype could be due to the disturbance of cellular signaling required for normal tip growth.

Compound C (benzovindiflupyr) has complex II of the respiratory chain as the target, and consequently an inhibition can lead to a depletion of energy. Therefore, depending on the concentration, the cell will stop growth because it lacks the ability to produce the required metabolites.

Compound X (a Syngenta research compound related to trifluoromethyloxadiazoles \cite{winter2020trifluoromethyloxadiazoles}, see Fig. \ref{fig: FigureS7} for the chemical structure) is an inhibitor of a histone deacetylase (HDAC). Histones are proteins that interact with nuclear DNA, and HDACs are enzymes that remove acetyl groups from histones. As a result, the DNA is packed more tightly, which has an influence on gene expression. HDAC-inhibitors result in de-regulation of this control mechanism for gene expression, however it is unknown why \textit{P. phakopsora} germlings react to this compound in the observed manner.

\section{Supplementary Methods}

Python was used for all computing, with the OpenCV library \cite{bradski2008learning} used for processing the snapshot images, and for data-driven development of the tip growth model. We manually built the PINN architecture using Tensorflow 2 \cite{abadi2016tensorflow}, however there is now a full PINN Python package, DeepXDE \cite{lu2021deepxde}. A Quadro RTX 6000 GPU card was used to speed up neural network training and inference. Finally, we used the pyABC library \cite{klinger2018pyabc} for parameter fitting of the tip growth model with ABC-SMC, the computationally efficient implementation of approximate Bayesian computation.

\section{Supplementary References}
\printbibliography[heading=none]